\documentclass{article}
\usepackage{graphicx}

\usepackage{hyperref}
\usepackage{amsmath}
\usepackage{amssymb}	
\usepackage{mathtools}

\usepackage{subcaption}
\usepackage[dvipsnames]{xcolor}

\usepackage{multirow}
\usepackage{pdfpages}
\usepackage{stfloats}
\usepackage{tabularx}
\usepackage[export]{adjustbox}
\usepackage{float}
\usepackage[
singlelinecheck=false 
]{caption}

\usepackage{multicol}

\begin{document}
\title{\bf On optical appearance of Einstein-Maxwell-{\AE}ther black holes surrounded by various accretions}
\author{{  Mitra Darvishi$^{1}$\thanks{\href{mailto:m-darvishi@pn.ac.ir}{m-darvishi@pn.ac.ir} }, Malihe Heydari-Fard$^{2}$%
\thanks{
Corresponding author: \href{mailto:m\_heydarifard@sbu.ac.ir}{heydarifard@qom.ac.ir}} and Morteza Mohseni$^{1}$\thanks{ \href{mailto:m-mohseni@pnu.ac.ir}{m-mohseni@pnu.ac.ir}}}\\{\small \emph{$^{1}$ Physics Department, Payame Noor University, Tehran 19395-3697, Iran } }
\\{\small \emph{$^{2}$ Department of Physics, The University of Qom, 3716146611, Qom, Iran}}}

\maketitle

\begin{abstract}
In this paper, we investigate the effects of the {\ae}ther field and the electric charge on the observed shadow of two types of charged black holes in the Einstein-Maxwell-{\AE}ther theory. By considering that the Einstein-Maxwell-{\AE}ther black holes surrounded by the static/infalling spherical accretion flows, as well as an optically and geometrically thin disk accretion flow, we study the shadow luminosities and the observed specific intensity of the image for these various profiles of accretion flows. We find that in the thin disk accretion model the location and the emitted model of the accretion gas affect on the optical appearance of charged Einstein-{\AE}ther black holes in contrast to the spherical accretion flows. For a thin disk profile, we show that the observer will receive more intensity for an emitted model as Gaussian function when the innermost radiation radius lies in the innermost stable circular orbit. Finally, comparing the results of the charged Einstein-{\AE}ther black holes with neutral Einstein-{\AE}ther black holes, we show that the charged Einstein-{\AE}ther black holes have smaller dark area, whereas wider lensed ring and photon ring. Comparing the charged Einstein-{\AE}ther black holes with Reissner-Nordstrom black hole gives a reverse effect.
\vspace{5mm}
\end{abstract}

\section{Introduction\label{section1}}
 The Event Horizon Telescope (EHT) as the very long baseline interferometry (VLBI) announced the first image of the supermassive black hole (BH) at the center of elliptical M87 galaxy \cite{A1}-\cite{A6}. Very recently it also released image of Sgr A* BH at the center of Milky Way as a spiral galaxy \cite{Sgr}. So observational evidence indicates that a supermassive BH resides in the center of many galaxies.

The first study of light deflection around a gravitationally intense star was carried out by Synge \cite{Synge}. Then Bardeen calculated the shadow radius of the Kerr BH and found that the rotation parameter caused that the shape of shadow is not a perfect circle \cite{Bardeen}. Since the astrophysical BHs surrounded by an extremely accretion flow, Luminet extended the synge's paper to a realistic BH \cite{Luminet}. He considered the Schwarzschild BH surrounded by a geometrically thin and optically thick accretion disk and showed that both the location and the specific emission of the accretion gas affect the optical appearance of BH and photon rings. By assuming a geometrically and optically thin accretion disk, Gralla et al. showed that depending on the intersection points between the light rays and the disk, the regions are separated the direct emission, lensed rings and photon rings. The results indicate that the direct emission has the dominant contribution to the optical appearance of the BH image \cite{Wald}. Also, for a Schwarzschild BH surrounded by static and infalling spherical accretion flow, Narayan et al. showed that in contrast to the thin disk model the outer radius of the BH shadow does not depend on the location of the inner edge of the radiating gas \cite{Narayan}. In recent years a lot of research about the BH shadow with and without accretion flow, including Schwarzschild BH surrounded by a quintessence matter \cite{Zeng1}, BH in Einstein-Gauss-Bonnet gravity \cite{Zeng2}, BH in Einstein-Dilaton-Gauss-Bonnet gravity \cite{000}, BH in Einstein-Maxwell-Chern-Simons gravity \cite{00}, BH in Einstein-Maxwell-Dilaton gravity \cite{Heydari-Fard:2021pjc}, Hayward BH \cite{0, 1}, Kehagias-Sfetsos BH \cite{ks2}, BH surrounded by a Bach Weyl ring \cite{2}, dual-charged String BH \cite{Stringy}, Kerr-Melvin BH \cite{4}, BH surrounded by cold dark matter halo \cite{5}, BH in Rastall gravity \cite{6, 7}, non-linear electrodynamic BH\cite{8, 9}, wormhole space-time \cite{10}, BH in Brane-world gravity \cite{11}, BH in alternative gravity theories \cite{12}, hairy BH \cite{13}, Bardeen BH  \cite{14}, regular BH \cite{15}, naked singularities \cite{16}, BH surrounded by a perfect fluid dark matter \cite{17, 18}, charged rotating BH in $f(R)$ gravity \cite{19} and Schwarzschild-MOG BHs in scalar-tensor-vector gravity \cite{MOG} have been extensively studied.

One of the fundamental principle of general relativity (GR) is the Lorentz invariance. But in high energy regime the local invariance may be violated \cite{Ahmadi:2006cr}--\cite{Bhattacharjee:2018nus}. Therefore a lot of fundamental gravity theories such as the Einstein-{\AE}ther theory and the Horava-Lifshitz theory of quantum gravity have been proposed as the Lorentz-violating gravitational theory. Einstein-{\AE}ther theory includes apart from the metric, a dynamical unit time-like vector field, $u^{\nu}$, that it defines a preferred time-like direction that unlike the GR violates the Lorentzian symmetry. This vector field is referred as the {\ae}ther field breaks the boost symmetry locally but not the rotation symmetries \cite{Jacobson:2000xp, Eling}. In the Einstein-{\AE}ther theory, the static and spherically symmetric charged BH solutions have been obtained \cite{Konoplya:2006ar}--\cite{Barausse:2013nwa}. Also, the BH solutions in the framework of the Einstein-Maxwell-{\AE}ther theory have been derived in Ref. \cite{Ding:2015kba, Ding:2016wcf}. For the spherically symmetric BH solutions in the Einstein-{\AE}ther theory by using numerical calculations see Ref. \cite{Eling:2006ec}. Slowly rotating BH solutions and their properties in the Einstein-{\AE}ther theory were presented in Ref. \cite{Barausse:2015frm}--\cite{Adam:2021vsk}. Authors in Ref. \cite{Zhu:2019ura} studied the BH shadow and the deflection angle of the charged rotating BHs in Einstein-{\AE}ther theory. Optical appearance of two types of the Einstein-{\AE}ther BHs surrounded by thin disk has been investigated by wang et al. \cite{Aether}. For the study of the quasi-normal modes in this theory  see Ref. \cite{Konoplya:2006rv}--\cite{Churilova:2020bql}. Also, the theoretical and observational constraints on the coupling constants of the Einstein-{\AE}ther theory by considering events GW 170817 and GRB 170817A and LIGO/Virgo data were studied in \cite{Oost:2018tcv, Alvarenga:2023eah} and \cite{Schumacher:2023cxh}, respectively.

As mentioned above in Ref. \cite{Aether} authors studied the optical appearance of the neutral Einstein-{\AE}ther BHs surrounded only by optically and geometrically thin accretion disk. In the present paper, we investigate the optical appearance of the charged Einstein-{\AE}ther BHs surrounded by two types of accretion flow, the static/infalling spherical accretion and thin disk accretion flow. The paper is organized as follows: In Sec. \ref{section2}, we give a brief review of Einstein-{\AE}ther theory. In Sec. \ref{section3}, we present the static charged BHs in Einstein-{\AE}ther gravity theory. The photon trajectories in the space-time of the charged Einstein-{\AE}ther BHs are investigate in Sec. \ref{section4}. In Sec. \ref{section5}, we present the shadow images of the charged Einstein-{\AE}ther BHs for both accretion flow static/infalling spherical and thin disk accretion. The results are summarized in Sec. \ref{section6}.

\section{Einstein-Maxwell-{\AE}ther theory\label{section2}}

Let us briefly review the Einstein-Maxwell-{\AE}ther theory of gravity. The action of the theory is defined
as
\begin{equation}
{\cal{ S}}=\int d^4x \sqrt{-g}\left[\frac{1}{16\pi G_{{\ae}}}\left(R+{\cal L}_{{\ae}}\right)+{\cal L}_{M}\right],
\label{1}
\end{equation}
where $g$ and $R$ are the determinant of the metric and the Ricci scalar, respectively. The {\ae}ther gravitational constant $G_{{\ae}}$ is related to the Newton's gravitational constant $G_{N}$ as
\begin{equation}
{G_{\ae}} = \frac{2 G_{N}}{2-c_{14}},
\label{2}
\end{equation}
where $c_{14} \equiv c_{1}+c_{4}$ is a coupling constant. The {\ae}ther Lagrangian is defined as
\begin{equation}
{\cal{ L_{{\ae}}}}\equiv-Z^{\alpha\beta}_{\,\,\,\,\,\,\,\mu\nu} \left(\nabla_{\alpha}u^{\mu}\right)\left(\nabla_{\beta}u^{\nu}\right)+\lambda\left(g_{\mu\nu}u^{\mu}u^{\nu}+1\right),
\label{3}
\end{equation}
where
\begin{equation}
Z^{\alpha\beta}_{\,\,\,\,\,\,\,\mu\nu} = c_{1} g^{\alpha\beta} g_{\mu\nu}+c_{2}\delta^{\alpha}_{\mu} \delta^{\beta}_{\nu}+c_{3}\delta^{\alpha}_{\nu}\delta^{\beta}_{\mu}-c_{4}u^{\alpha}u^{\beta}g_{\mu\nu}.
\label{4}
\end{equation}
Here, $c_{i}(i=1,2,3,4)$ are coupling constants of the theory, $\lambda$ is lagrangian multiplier and $u^{2}=g_{\mu\nu}u^{\mu}u^{\nu}$ with $u^{\mu}$ is a time-like four vector and unity guaranteed by $\lambda$. The Maxwell lagrangian is
\begin{equation}
{\cal{L}}_{M}=-\frac{1}{16\pi G_{{\ae}}}F_{\mu\nu}F^{\mu\nu},
\label{5}
\end{equation}
with
\begin{equation}
F_{\mu\nu}=\partial_{\mu}A_{\nu}-\partial_{\nu}A_{\mu},
\label{6}
\end{equation}
where $A_{\mu}$ is the electromagnetic vector potential. This electromagnetic field is minimally coupled to both the gravity $g_{\mu\nu}$ and the {\ae}ther field $u^{\mu}$.

By varing the action (\ref{1}) with respect to $g_{\mu\nu},u^{\mu}$, $\lambda$ and $A_{\mu}$, we obtain the following field equations
\begin{equation}
R^{\mu\nu}-\frac{1}{2}g^{\mu\nu}R-8\pi G_{æ}T{^{\mu\nu}_{æ}}=0,
\label{7}
\end{equation}

\begin{equation}
\nabla_{\mu}J^{\mu}_{\alpha}+c_{4}a_{\mu}\nabla_{\alpha}u^{\mu}+\lambda u_{\alpha} = 0,
\label{8}
\end{equation}

\begin{equation}
g_{\mu\nu}u^{\mu}u^{\nu}=-1,
\label{9}
\end{equation}

\begin{equation}
\nabla^{\mu}F_{\mu\nu}=0,
\label{10}
\end{equation}
with
\begin{eqnarray}
T^{{\ae}}_{\alpha\beta}  &\equiv& \nabla_{\mu}\left[J ^{\mu}_{(\alpha}u_{\beta)}+J_{(\alpha\beta)}u^{\mu}-u_{(\beta}J^{\mu}_{\alpha)}\right]
+c_{1}\left[(\nabla_{\alpha} u_{\mu}) (\nabla_{\beta}u^{\mu})-(\nabla_{\mu}u_{\alpha})(\nabla^{\mu}u_{\beta})\right] \nonumber\\
&+&c_{4}a_{\alpha} a_{\beta}+\lambda u_{\alpha} u_{\beta}-\frac{1}{2}g_{\alpha\beta} J^{\delta}_{\sigma}\nabla_{\delta}u^{\sigma},
\label{11}
\end{eqnarray}
and
\begin{equation}
J^{\alpha}_{\mu}\equiv Z^{\alpha\beta}_{\,\,\,\,\,\,\,\,\mu\nu} \nabla_{\beta}u^{\nu},
\label{12}
\end{equation}
\begin{equation}
a^{\mu}\equiv u^{\alpha}\nabla_{\alpha}u^{\mu}.
\label{13}
\end{equation}
In Ref. \cite{Jacobson} authors have been obtained a number of constraints on the coupling constants by combining the theoretical and observational results. In this paper in order to investigate the effects of these coupling constants and charge parameter on the shadow size, photon sphere and photon rings, we choose the following constraints
\begin{equation}
0\leqslant c_{13}<1,\hspace {0.5 cm} 0\leqslant c_{14}<2,\hspace {0.5 cm} 2+c_{13}+3c_{2}>0.
\label{14}
\end{equation}

\section{Charged BHs in Einstein-Maxwell-{\AE}ther theory\label{section3}}
The static spherically symmetric space-time for the Einstein-Maxwell-{\AE}ther theory can be written as
\begin{equation}
ds^{2}=-f(r)dt^{2}+f^{-1}(r)dr^{2}+r^{2}\left(d\theta^{2}+\sin^{2}\theta d\varphi^{2}\right).
\label{010}
\end{equation}
Depending on the choice of coupling constants there are two types of solutions corresponding to $c_{123}\neq 0$ and $c_{123}=0$ \cite{Ding:2015kba},

$\bullet$ {\textbf{First type}} $c_{123} \neq 0$ :

For the first type of the Einstein-Maxwell-{\AE}ther BH, we have
\begin{equation}
f_{1}(r)=1-\frac{2M}{r}+\frac{q^{2}}{r^{2}}+\frac{c_{13}B}{\left(1-c_{13}\right)r^{4}},
\label{012}
\end{equation}
where $M$ is the BH mass, $q$ is the BH charge and $B$ is defined as
\begin{equation}
B=\frac{\left(2M-\sqrt{-32q^{2}+36M^{2}}\right)\left(6M+\sqrt{-32q^{2}+36M^{2}}\right)^{3}}{4096}.
\label{13}
\end{equation}
Obviously, one can obtain the Schwarzschild BH solution when $c_{13}=q=0$ . For $c_{13}=0$, $q\neq0$ and $c_{13}\neq0$, $q=0$ the above solution reduces to Reissner-Nordstrom (RN) BH and the first neutral Einstein-{\AE}ther BH solutions, respectively. The event horizon $r_h$ can be obtained by solving $f_{1}(r)=0$, which for this type has a complicated form and therefore we will not bring it here.

$\bullet$ {\textbf{Second type}} $c_{123} = 0$ :

The metric function for the second type of the charged Einstein-{\AE}ther BH is given by
\begin{equation}
f_{2}(r)=1-\frac{2M}{r}-\frac{r_{u}\left(2M+r_{u}\right)}{r^{2}},
\end{equation}
where
\begin{equation}
r_{u} = M \left[\sqrt{\frac{\left(2-c_{14}\right)}{2\left(1-c_{13}\right)}-\frac{q^{2}}{M^{2}\left(1-c_{13}\right)}}-1\right].
\label{14}
\end{equation}
Since the expression under square root in Eq.~(\ref{14}) should be always positive, we obtain the following constraints
\begin{equation}
q\leq M\sqrt{c_{13}-\frac{c_{14}}{2}},\hspace {0.5 cm}c_{13}\geq\frac{c_{14}}{2}.
\label{15}
\end{equation}
By solving $f_2(r)=0$, we obtain the event horizon of the second charged Einstein-{\AE}ther BH solution as
\begin{equation}
r_{h}=M\left[1+\sqrt{\frac{c_{14}-2+\frac{2q^{2}}{M^{2}}}{2\left(c_{13}-1\right)}}\right],
\label{16}
\end{equation}
where for $q=0$ reduces to equation (20) from Ref. \cite{Aether}. For $c_{13}=c_{14}=q=0$ and $c_{13}=c_{14}=0$, we recover the schwarzschild event horizon, $r_{h}=2M$, and the RN event horizon, $r_{h}=M+\sqrt{M^2-q^2}$, respectively. In what follows, we consider two specific cases of the second charged Einstein-{\AE}ther BHs; when $c_{14} = 0$ and when $c_{13} = 0$.

\section{Null geodesic in Einstein-Maxwell-{\AE}ther space-time\label{section4}}
The Lagrangian for a test particle in the charged Einstein-{\AE}ther BH is given by
\begin{equation}
{\cal{L}}=\frac{1}{2}g_{\mu\nu}\dot{x}^{\mu}\dot{x}^{\nu}=\frac{1}{2}\left[-f(r) \dot{t}^{2}+{f(r)}^{-1}{\dot{r}^{2}}+r^{2}\left(\dot{\theta}^{2}+\sin^{2}\theta \dot{\varphi}^{2}\right)\right],
\label{17}
\end{equation}
where $\dot{x^{\mu}}$ is the four-velocity of light and a dot presents the derivative with respect to the affine parameter $\tau$. For $\theta=\frac{\pi}{2}$ and $\dot{\theta}=0$ in the equatorial plane, the Euler-Lagrange equations for $t$ and $\varphi$ coordinates can be obtained as
\begin{equation}
{{E}}=-\frac{\partial l}{\partial \dot{t}}=f(r)\dot{t},
\label{18}
\end{equation}
\begin{equation}
{\cal{L}}=\frac{\partial l}{\partial\dot{\varphi}}=r^{2}\dot{\varphi}.
\label{19}
\end{equation}
Now, for null geodesics with condition ${\cal L}=0$, we find
\begin{equation}
k^{t}\equiv\dot{t}=\frac{1}{bf(r)},
\label{20}
\end{equation}
\begin{equation}
k^{\varphi}\equiv\dot{\varphi}=\frac{1}{r^{2}},
\label{21}
\end{equation}
\begin{equation}
k^{r}\equiv\dot{r}=\sqrt{\frac{1}{b^2}-V_{eff}(r)},
\label{22}
\end{equation}
where $b\equiv\frac{L}{E}$ is the impact parameter and the effective potential for the charged BH in Einstein-Maxwell-{\AE}ther theory is defined as
\begin{equation}
V_{\rm eff}=\frac{f(r)}{r^{2}},
\label{23}
\end{equation}
here $f_{1}(r)$ and $f_{2}(r)$ denote the effective potential for the first and second charged Einstein-{\AE}ther BH, respectively.

\begin{figure}[H]
\centering
\includegraphics[width=3.0in]{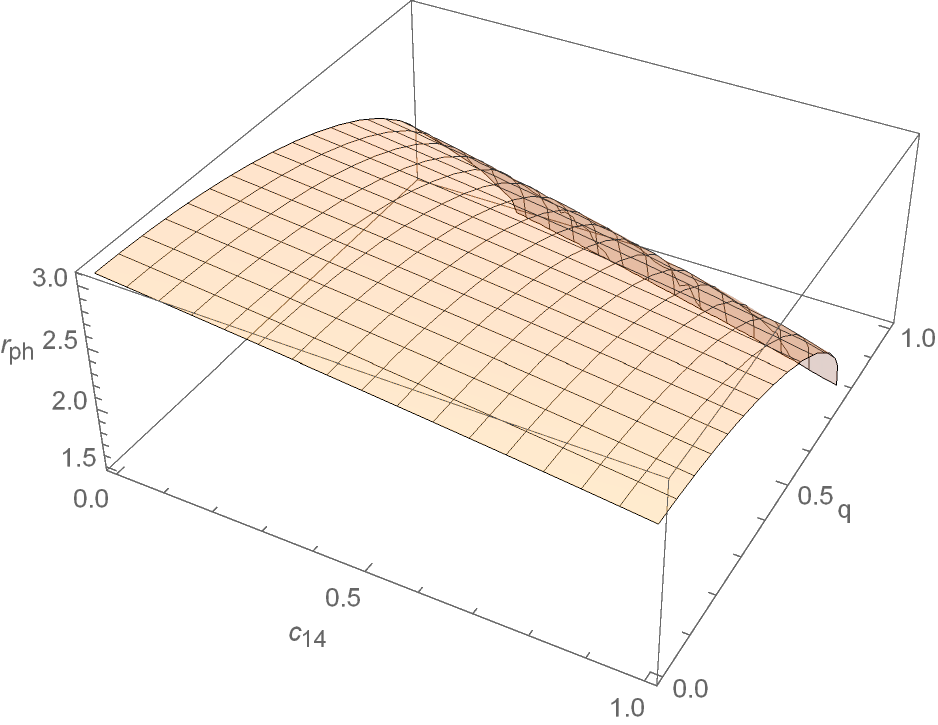}
\includegraphics[width=3.0in]{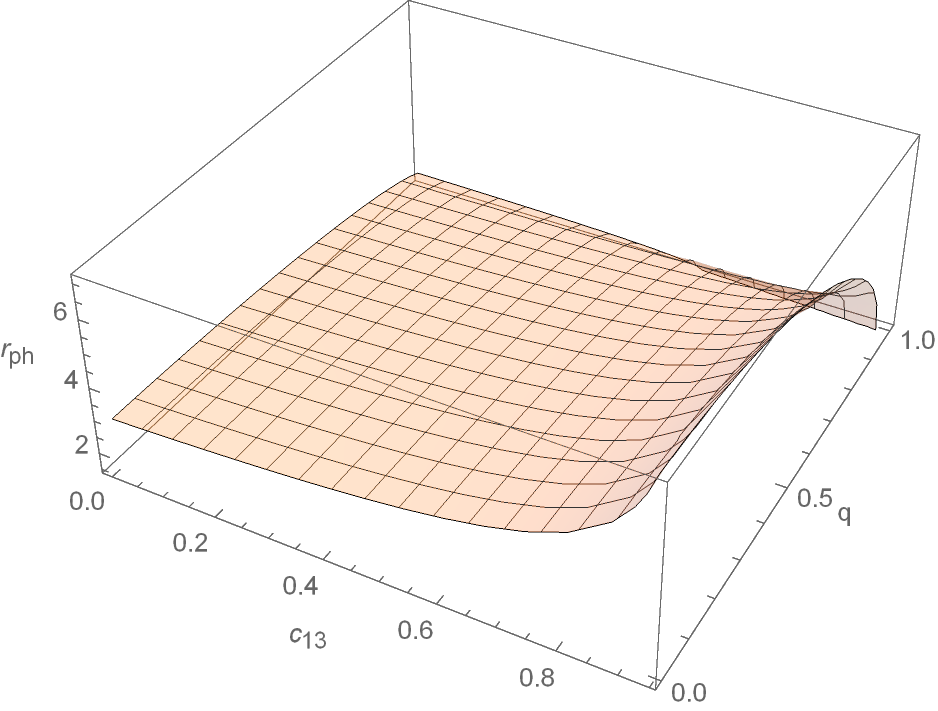}
\caption{ The behavior of the photon sphere for the second charged Einstein-{\AE}ther BHs, when $c_{13} = 0$ for different value of $c_{14}$ (left panel) and when $c_{14} = 0$ for different value of $c_{13}$ (right panel) with $0\leq q \leq 1$ and $M = 1$.}
\label{3D}
\end{figure}

For the photon orbit, $r_{\rm ph}$, with $r =  r_{\rm ph} = {\rm const}$, we obtain two following constraints from Eq.(\ref{22})
\begin{equation}
V_{\rm eff}(r_{\rm ph})=\frac{1}{b^{2}},
\label{24}
\end{equation}
\begin{equation}
{V^{'}_{\rm eff}}(r_{\rm ph})=0,
\label{25}
\end{equation}
where the prime presents the derivative with respect to variable $r$. For the first Einstein-Maxwell-{\AE}ther BH two quantities photon sphere $r_{ph}$ and the critical impact parameter $b_{ph}$ have complicated forms, while for the second Einstein-Maxwell-{\AE}ther BH, from Eqs.~(\ref{24}) and (\ref{25}) we have
\begin{equation}
r_{\rm ph}=\frac{3M}{2}\left[1+\sqrt{1-\frac{4c_{14}+8\frac{q^{2}}{M^{2}}-8c_{13}}{9\left(1-c_{13}\right)}}\right].
\label{26}
\end{equation}
Clearly, the above equation for $q=0$ reduces to relation (28 a) from Ref. \cite{Aether}. Also, for $c_{13}=c_{14}=q=0$ and $c_{13}=c_{14}=0$ it reduces to the photon sphere of the Schwarzschild BH $r_{\rm ph}=3M$ and the photon sphere of Reissner-Nordstrom (RN) BH
\begin{equation}
r_{ph}=\frac{3M}{2}\left[1+\sqrt{1-\frac{8q^{2}}{9m^{2}}}\right],
\label{26a}
\end{equation}
respectively. In Fig. \ref{3D}, we have plotted the behavior of the photon sphere, $r_{\rm ph}$, of the second charged Einstein-{\AE}ther BHs for two cases when $c_{14} = 0$ and when $c_{13} = 0$.

Now, by substituting $r_{\rm ph}$ in Eq.~(\ref{24}) one can find the following critical impact parameter as
\begin{equation}
b_{\rm ph}=\frac{r_{ph}}{\sqrt{f(r_{\rm ph})}}=
\frac{ 3(-1+c_{13})M+ \sqrt{-1+c_{13}} \sqrt{(-9+c_{13}+4c_{14})M^2+8q^2} } {(-1+c_{13}) \sqrt{
\frac{(3+c_{13}-2c_{14})-4 q^2+M \sqrt{-1+c_{13}} \sqrt{(-9+c_{13}+4c_{14})M^2+8q^2}}{(2c_{13}-c_{14})M^2-2q^2}}}.
\label{27}
\end{equation}

The results of the BH event horizon $r_{\rm h}$, photon sphere $r_{\rm ph}$ and impact parameter $b_{\rm ph}$ for the first charged Einstein-{\AE}ther BH are presented in Tab. \ref{table1} . $r_{\rm h}$, $r_{\rm ph}$ and $b_{\rm ph}$ increase by increasing the coupling constant $c_{13}$, whereas the opposite effect occurs by increasing the charge parameter. For the second charged Einstein-{\AE}ther BH we take into account two cases; when $c_{14} = 0$ and when $c_{13} = 0$. For case $c_{14} = 0$ of the second charged Einstein-{\AE}ther BH the result is similar to the first charged Einstein-{\AE}ther BH. For case $c_{13} = 0$, $r_{\rm h}$, $r_{\rm ph}$ and $b_{\rm ph}$ decrease by increasing both the coupling constant and charge parameter. For two cases of the second charged Einstein-{\AE}ther BH the results are summarized in Tab. \ref{table2} and \ref{table3}.

Now, we concentrate on the light trajectories around charged Einstein-{\AE}ther BHs with different coupling constants and charge parameter. From Eqs.~(\ref{21}) and (\ref{22}), we have the following equation of motion
\begin{equation}
\frac{dr}{d\phi}=\pm r^{2}\sqrt{\frac{1}{b^{2}}-V_{\rm eff}(r)},
\label{28}
\end{equation}
where the effective potential for the first and second types solutions are defined as
\begin{equation}
V^{(1)}_{\rm eff}(r)=\frac{1}{r^{2}}\left[1-\frac{2M}{r}+\frac{Q^{2}}{r^{2}}+\frac{c_{13}B}{\left(1-c_{13}\right)} \frac{1}{r^{4}}\right],
\label{29}
\end{equation}
and
\begin{equation}
 V^{(2)}_{\rm eff}(r)=\frac{1}{r^{2}}\left[1-\frac{2M}{r}-\frac{r_{u}\left(2M+r_{u}\right)}{r^{2}}\right],
\label{30}
\end{equation}
that $B$ and $r_u$ were defined in Eqs.~(\ref{13}) and (\ref{14}). Note, from Eq.~(\ref{28}) the light trajectories are determined by impact parameter, charge and coupling constants.

\begin{table}[H]
\centering
\caption{\footnotesize  The values of photon radius, $r_{\rm ph}$, impact parameter, $b_{\rm ph}$, and  the event horizon radius $r_{\rm h}$ of the first charged Einstein-{\AE}ther BH for different values of $q$ and $c_{13}$.}
\begin{tabular}{l l l l l l l l}
\hline
Type&$q$&$c_{13}$& $r_{\rm h}/M$&$r_{\rm ph}/M$& $b_{\rm ph}/M$\\ [0.5ex]
\hline
Schwarzschild &0  &0          & 2 & 3&5.19615\\
Reissner-Nordstrom &0.5  &  0      &1.86603 &2.82288 &4.96791\\
Neutral-Einstein-{\AE}ther &0  &0.7       &2.31668 &3.32216 &5.49394\\
Charged-Einstein-{\AE}ther &0.3  &0.7       & 2.26085 &3.24862 &5.40173\\
Charged-Einstein-{\AE}ther &0.5  &0.7       &2.15403 &3.10872 &5.22777\\
\hline
\label{table1}
\end{tabular}
\end{table}

\begin{table}[H]
\centering
\caption{\footnotesize  The values of photon radius, $r_{\rm ph}$, impact parameter, $b_{\rm ph}$, and  the event horizon radius $r_{\rm h}$ of the second charged Einstein-{\AE}ther BH when $c_{14} = 0$ for different values of $c_{13}$, and $q$.}
\begin{tabular}{l l l l l l l l}
\hline
Type&$q$&$c_{13}$&$c_{14}$& $r_{\rm h}/M$&$r_{\rm ph}/M$&$b_{\rm ph}/M$\\ [0.5ex]
\hline
 Schwarzschild &    0 &0             &0  &2  &3 &5.19615\\
 Reissner-Nordstrom &    0.5   & 0         &0   &1.86603   &2.82288&4.96791 \\
 Neutral-Einstein-{\AE}ther &    0    &0.7           &0  &2.82574   &4.12996&6.70910 \\
 Charged-Einstein-{\AE}ther &    0.3 &0.7            &0  &2.74165   &4.01330&6.55007 \\
 Charged-Einstein-{\AE}ther &  0.5  &0.7           &0  &2.58114   &3.79129&6.24875 \\
\hline
\label{table2}
\end{tabular}
\end{table}

\begin{table}[H]
\centering
\caption{\footnotesize  The values of photon radius, $r_{\rm h}$, impact parameter, $b_{\rm h}$ and  the event horizon radius $r_{\rm h}$ of the second charged Einstein-{\AE}ther BH when  $c_{13} = 0$ for different values of $c_{14}$ and $q$.}
\begin{tabular}{l l l l l l l l}
\hline
Type&$q$&$c_{13}$&$c_{14}$& $r_{\rm h}/M$&$r_{\rm ph}/M$&$b_{\rm ph}/M$\\ [0.5ex]
\hline
 Schwarzschild &    0 &0             &0  &2  &3 &5.19615\\
 Reissner-Nordstrom &    0.5   & 0         &0   &1.86603   &2.82288&4.96791 \\
 Neutral-Einstein-{\AE}ther &   0   & 0         &0.7   &1.80623   &2.74499&4.86889 \\
 Charged-Einstein-{\AE}ther &   0.3 &0            &0.7  &1.74833   &2.67047&4.77504 \\
 Charged-Einstein-{\AE}ther &  0.5  &0           &0.7  &1.63246  &2.52470& 4.59449\\
\hline
\label{table3}
\end{tabular}
\end{table}

\begin{figure}[H]
\centering
\includegraphics[width=3.5in]{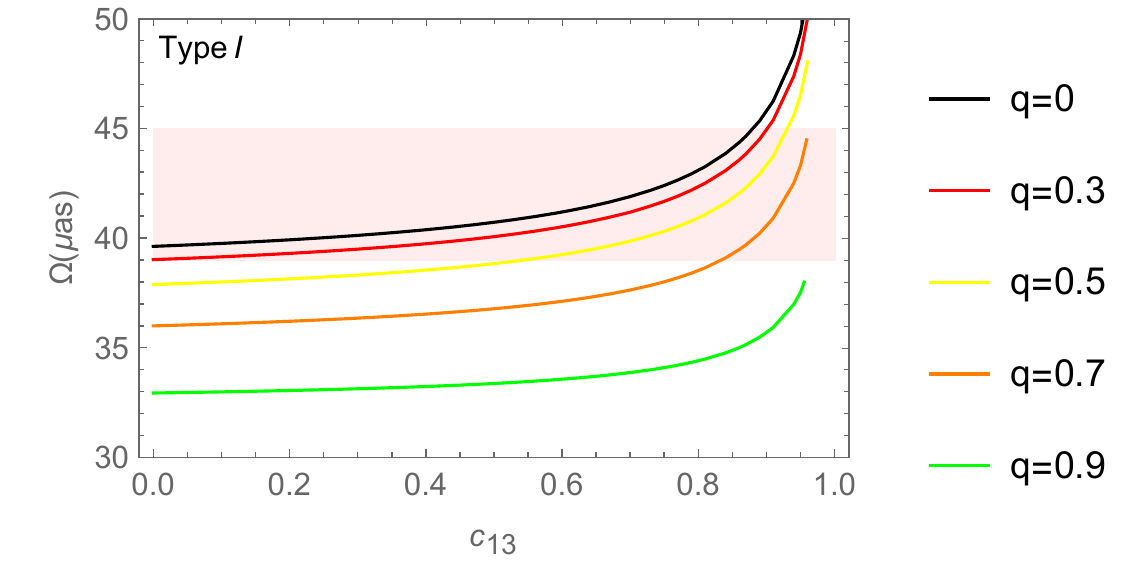}
\caption{\footnotesize Shadow diameter of the first charged Einstein-{\AE}her BH as a function of coupling constant $c_{13}$. The pink region shows the experimental data of the M87* BH ($42\pm3$ ${\rm \mu as}$) released by EHT.}
\label{constraint1}
\end{figure}

As often has been demonstrated the BH shadow radius depends on the geometry of the space-time, so the coupling constant of the Einstein-Maxwell-{\AE}ther theory can be constrained by the observed shadow. Considering a distant observer, the angular diameter $\Omega$ of the BH shadow can be written as \cite{Perlick:2021aok}
\begin{equation}
\left(\frac{\Omega}{\rm {\mu as}}\right)=\left(\frac{6.191165\times 10^{-8}}{\pi}\frac{\gamma}{\frac{D}{Mpc}}\right)\left(\frac{b_{\rm ph}}{M}\right),
\end{equation}
here $D$ is the distance between the distant observer and BH, $b_{\rm ph}$ is the critical impact parameter and $\gamma$ is the mass ratio of the BH to the Sun star. From the data of M87* BH, $D=16.8$ Mpc, $\gamma=6.5\times10^9$ and $\Omega= 42\pm3$ ${\rm \mu as}$ \cite{A1}--\cite{Sgr}. We plotted $\Omega$ as a function of $c_{13}$ with various of $q$ in Fig. \ref{constraint1} for the first charged Einstein-{\AE}her BH. As the figure shows, we constrain the parameter $0<c_{13}<0.88073$ for $q=0$ and $0<c_{13}<0.90205$ for $q=0.3$. We also obtain the constraint on the coupling constant $0<c_{13}<0.48535$ for $q=0$ and $0<c_{13}<0.53167$ for $q=0.3$ for the second charged Einstein-{\AE}her BH when $c_{14} = 0$ and $0<c_{14}<0.18419$ for $q=0$ and $0<c_{14}<0.16420$ for $q=0.3$ for the second charged Einstein-{\AE}her BH when $c_{13} = 0$.

\begin{figure}[H]
\centering
\includegraphics[width=3.1in]{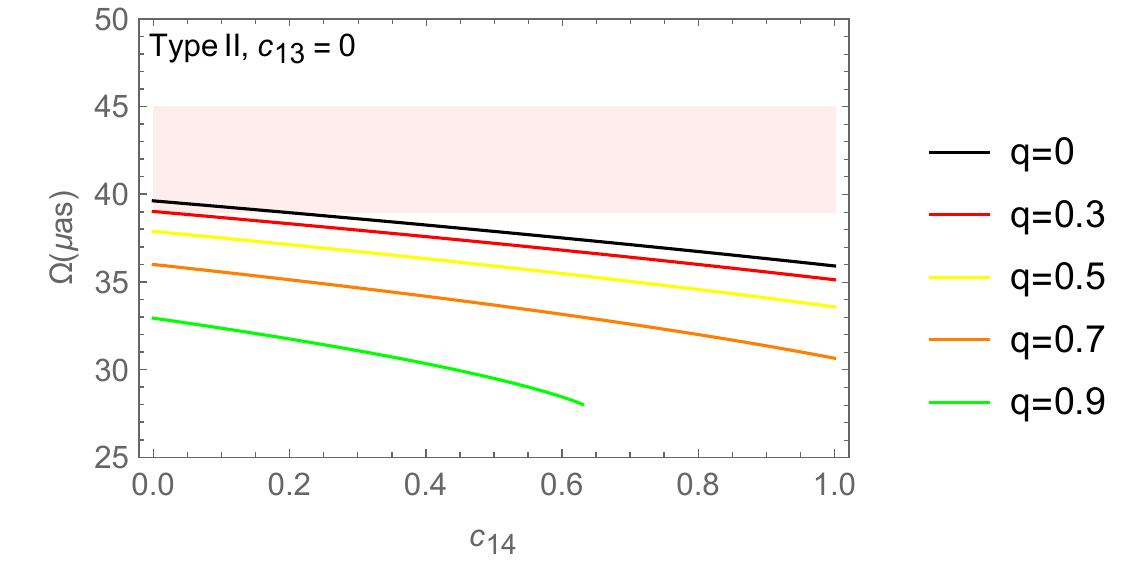}
\includegraphics[width=3.1in]{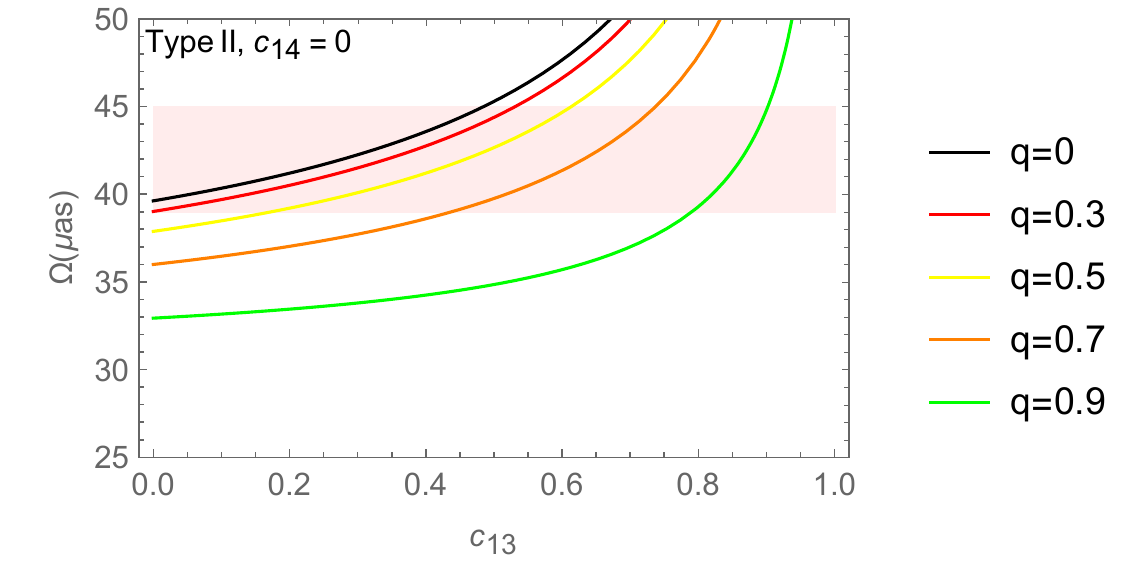}
\caption{\footnotesize Shadow diameter of the second charged Einstein-{\AE}her BH as a function of coupling constant $c_{14}$ with $c_{13} = 0$ (left panel) and as a function of coupling constant $c_{13}$ with $c_{14} = 0$ (right panel). The pink region shows the experimental data of the M87* BH ($42\pm3$ ${\rm \mu as}$) released by EHT.}
\label{constraint2}
\end{figure}

\section{Optical appearance of charged Einstein-{\AE}ther BH nuder three accretion flow models\label{section5}}
\label{1-Shadows and photon rings with spherical accretion flow}

\subsection{Static spherical accretion}
First, we study the shadow and photon rings for a charged Einstein-{\AE}ther BHs surrounded by static spherical accretion. We can obtain the observed specific intensity (${\rm erg s^{-1} cm^{-2} str^{-1} Hz^{-1}}$) by integrating the specific emissivity along any light ray path \cite{flow1, flow2}
\begin{equation}
I(\nu^{\rm s}_{\rm ob}) = \int_{\gamma} {g^{\rm s}}^3 j(\nu^{\rm s}_{\rm em}) dl_p,
\label{32}
\end{equation}
in which ${g^{\rm s}} \equiv \frac{{\nu^{\rm s}}_{\rm ob}}{{\nu^{\rm s}}_{\rm em}}$ is the redshift factor, ${\nu^{\rm s}}_{\rm ob}$ and ${\nu^{\rm s}}_{\rm em}$ denote the observed photon frequency and intrinsic photon frequency, respectively. For two types of charged Einstein-{\AE}ther BHs \cite{Ding:2015kba}, the redshift factor is defined as ${g^{\rm s}} = f(r)^{1/2}$. By assuming that the light radiation is monochromatic with rest frame frequency $\nu_{\rm t}$, the specific emissivity takes the form
\begin{equation}
j ({\nu^{\rm s}}_{\rm em}) \propto \frac{\delta({\nu^{\rm s}}_{\rm em}-\nu_{t})}{r^2},
\label{33}
\end{equation}
where we have taken the emission radial profile as $1/r^2$ \cite{flow2}. Also we can obtain the proper length measured in the rest frame of the emitter from Eq.~(\ref{010}) by the following relation
\begin{equation}
dl_{\rm p} = \sqrt{\frac{1}{f(r)}dr^2+r^2 d\varphi^2} = \sqrt{\frac{1}{f(r)}+r^2 \left(\frac{d\varphi}{dr}\right)^2} dr.
\label{34}
\end{equation}
Now, by substituting Eq.~(\ref{34}) into Eq.~(\ref{32}), the specific intensity observed by a distant observer is given by
\begin{equation}
I(\nu^{\rm s}_{\rm ob})=\int_{\gamma}\frac{f(r)}{r^{2}}\sqrt{1+\frac{b^{2}f(r)}{r^{2}-b^{2}f(r)}}dr.
\label{35}
\end{equation}

\begin{figure}[H]
\centering
\includegraphics[width=3.0in]{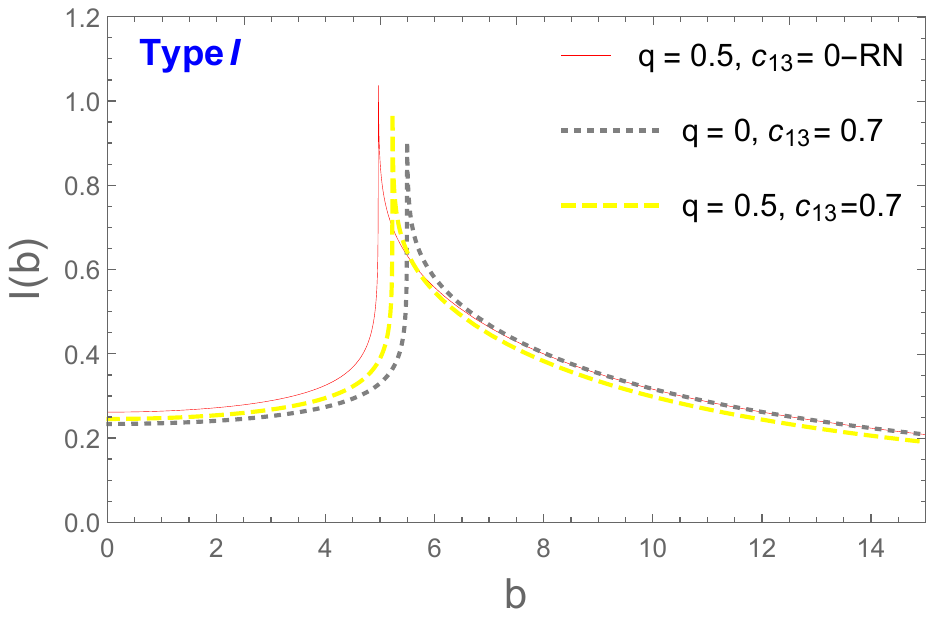}
\includegraphics[width=3.0in]{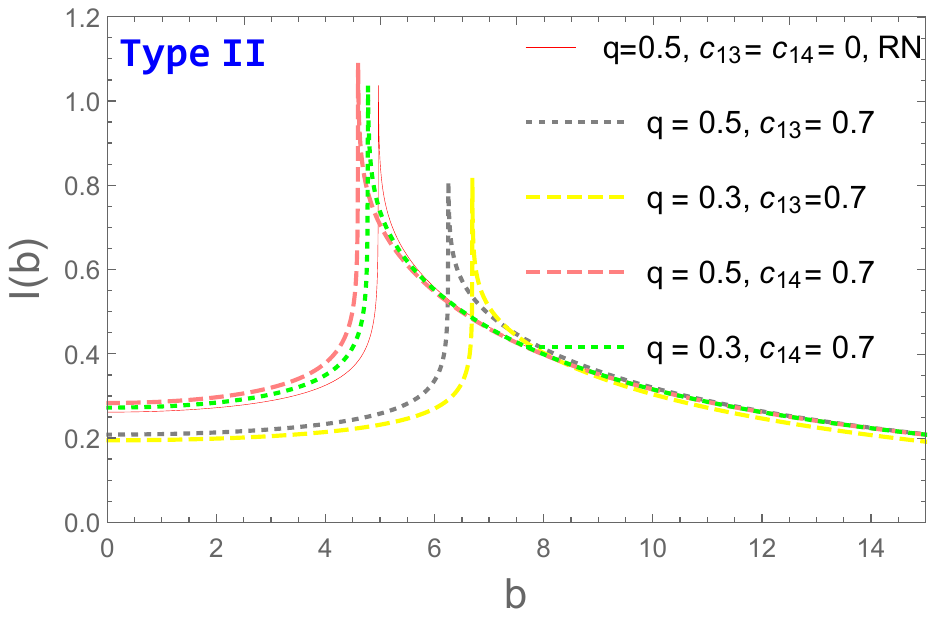}
\caption{ The observed intensity $I(\nu_{\rm ob})$ for static spherical accretion flow around the first charged Einstein-{\AE}ther BH (left panel) and the second charged Einstein-{\AE}ther BH (right panel). In each panel the red solid curve corresponds to the RN BH. }
\label{intensity-static}
\end{figure}

\begin{figure}[H]
\centering
\includegraphics[width=1.8in]{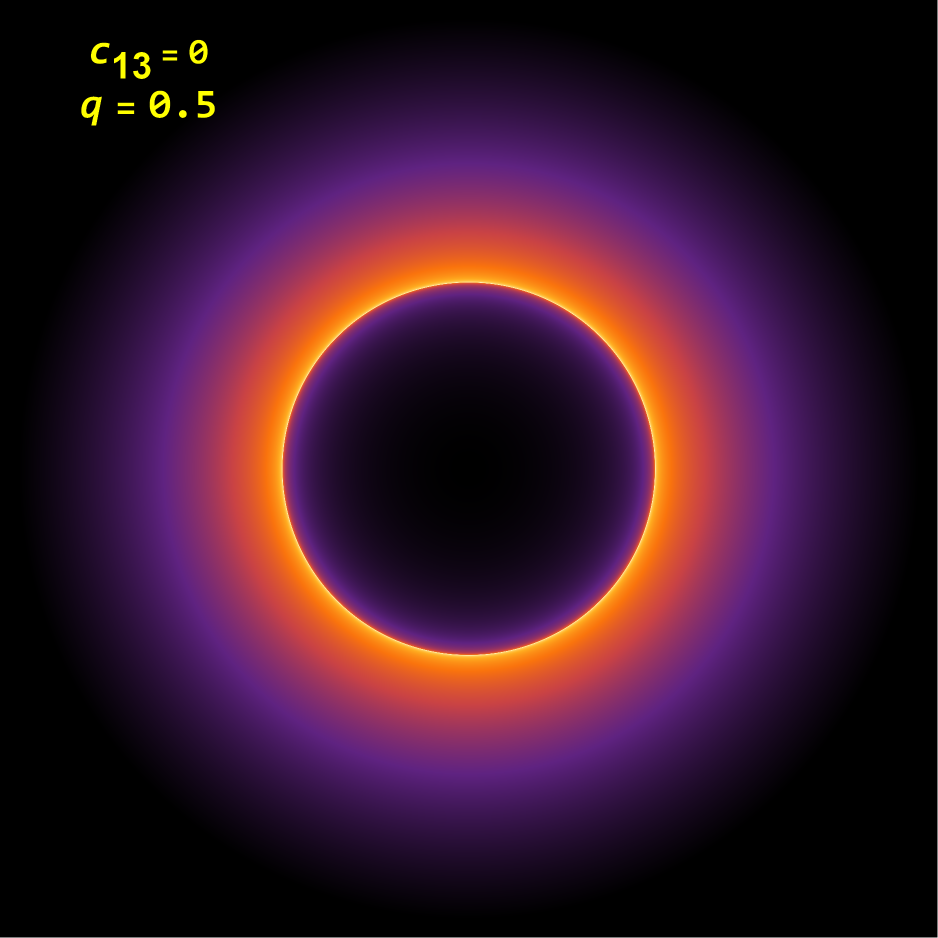}
\includegraphics[width=0.2in]{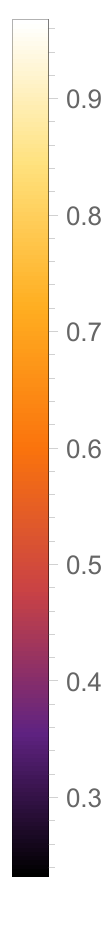}
\includegraphics[width=1.8in]{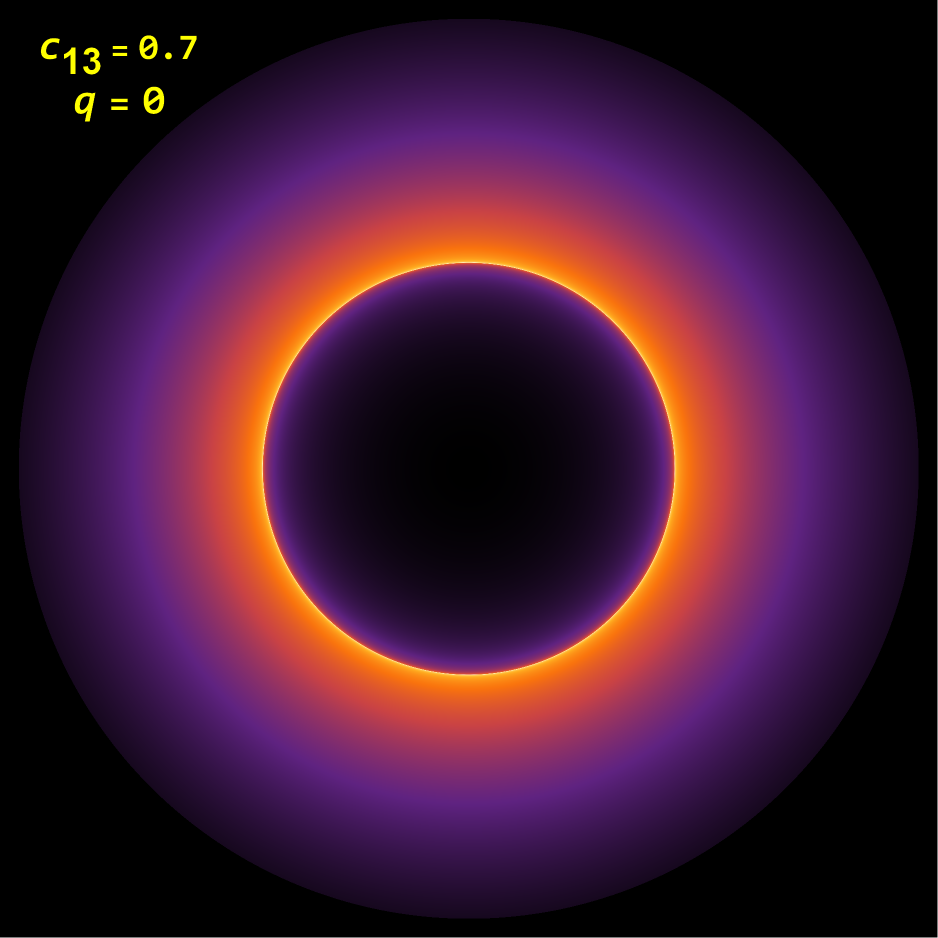}
\includegraphics[width=0.2in]{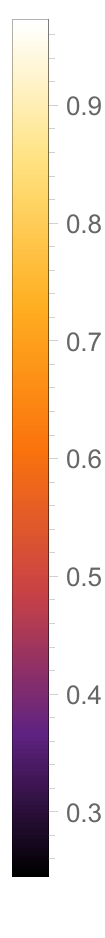}
\includegraphics[width=1.8in]{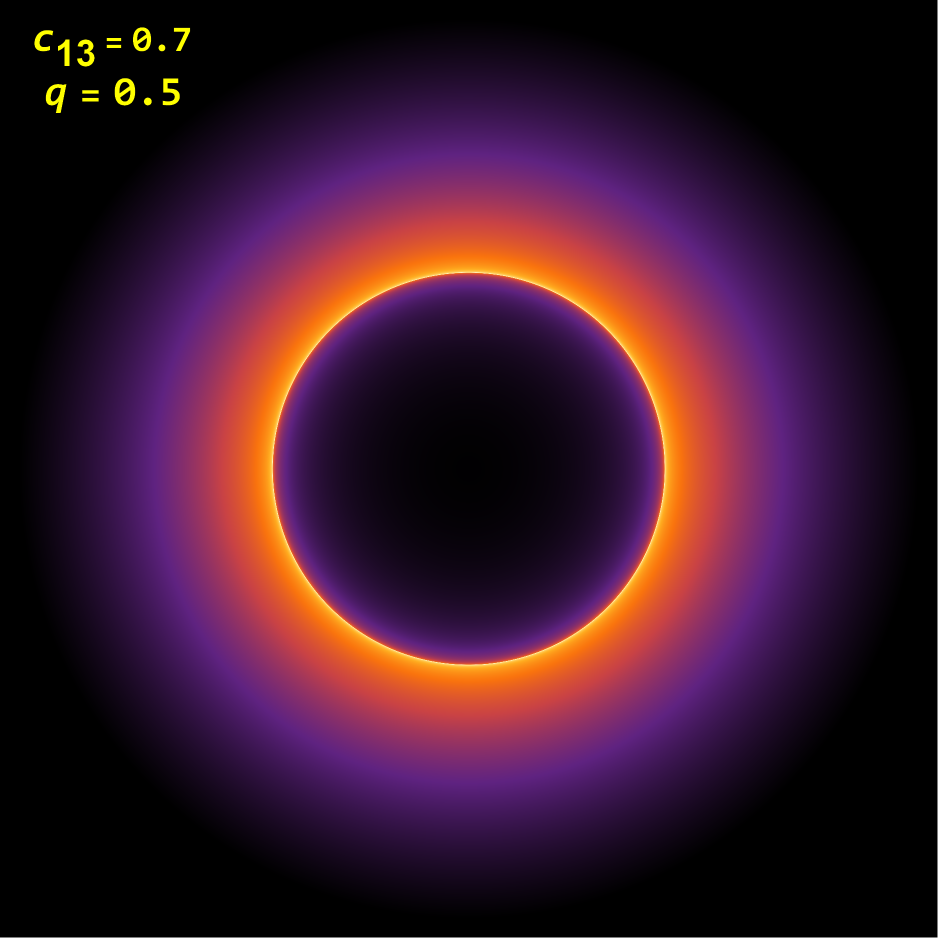}
\includegraphics[width=0.2in]{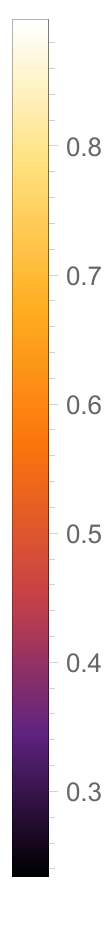}
\caption{Two dimensional image of shadows and photon rings for static spherical accretion flow of the RN BH with $c_{13}=0$, $q=0.5$ (left panel), the first neutral Einstein-{\AE}ther BH with $c_{13}=0.7$, $q=0$ (middle panel) and the first charged Einstein-{\AE}ther BH with $c_{13}=0.7$, $q=0.5$ (right panel) and $M = 1$.}
\label{2-dimensional-2static}
\end{figure}

The left panel of Fig. \ref {intensity-static} shows the observed specific intensity $I(\nu^{\rm s}_{\rm ob})$ as a function of the impact parameter $b$ for the first charged Einstein-{\AE}ther BH. The observed intensity increases with the impact parameter and reaches a maximum value at $b_{\rm ph}$ and then decreases with the impact parameter. The peak of intensity in $b_{\rm ph}$ due to the light rays revolve around the BH several times. The peak value of the observed intensity decreases with increasing the coupling constant $c_{13}$ when the charge parameter $q$ is constant and also it increases with increasing $q$ when $c_{13}$ is constant. Two dimensional image of shadows and photon rings for static spherical accretion of the first charged Einstein-{\AE}ther BH has been also plotted in Fig. \ref{2-dimensional-2static}. Clearly, for a fixed value of $c_{13}$ with increasing $q$ the luminosities of shadow and photon rings increase while for a fixed vale of $q$ with increasing $c_{13}$ the result is inverted. This is due to the fact that the gravitational field of the neutral Einstein-{\AE}ther BHs is stronger than the charged Einstein-{\AE}ther BHs which increases the deflection of light. This causes more photons to be trapped by the BH and a lower luminosities of shadow and photon rings being observed compared to the charged Einstein-{\AE}ther BHs.

Similarly, in the right panel of Fig. \ref {intensity-static}, we plotted the observed specific intensity for the second charged Einstein-{\AE}ther BH when $c_{14}=0$ for different values of $c_{13}$ and $q$ and also for the case when $c_{13}=0$ for different values of $c_{14}$ and $q$ with $M=1$.

\subsection{Infalling spherical accretion}
\begin{figure}[H]
\centering
\includegraphics[width=3.0in]{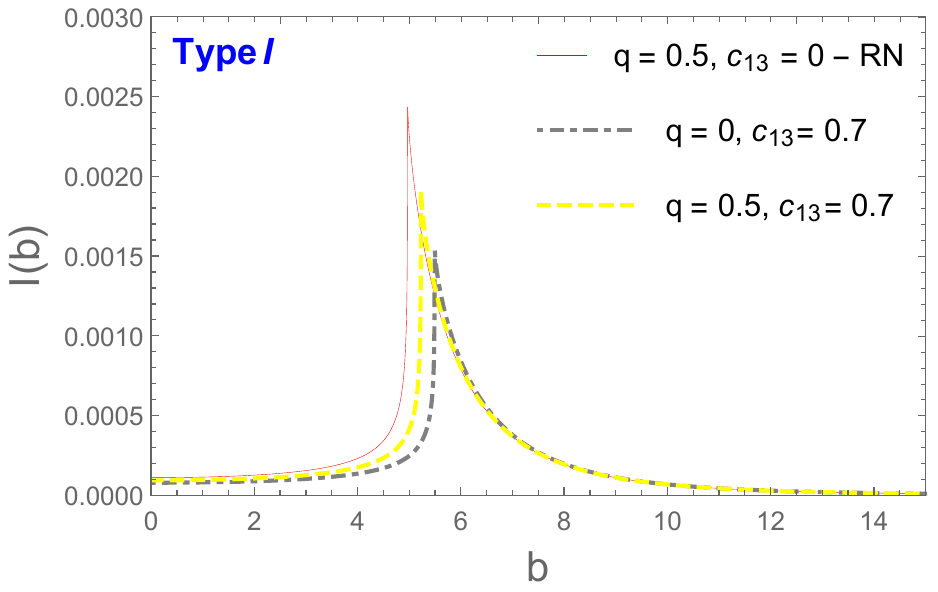}
\includegraphics[width=3.0in]{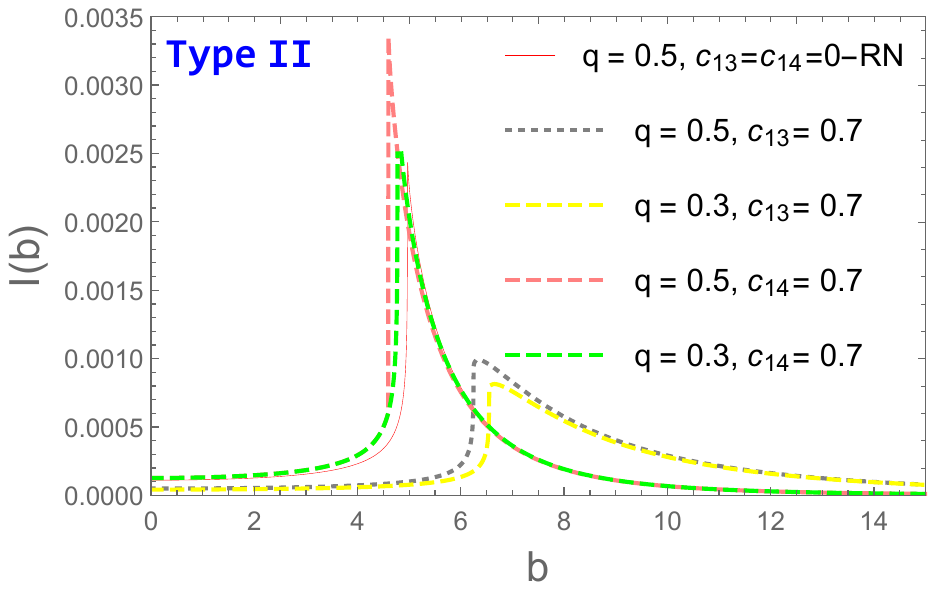}
\caption{ The observed intensity $I(\nu_{\rm ob})$ for infalling spherical accretion flow around the first charged Einstein-{\AE}ther BH (left panel) and the second charged Einstein-{\AE}ther BH (right panel). In each panel the red solid curve corresponds to the RN BH. }
\label{intensity-infalling}
\end{figure}

\begin{figure}[H]
\centering
\includegraphics[width=1.7in]{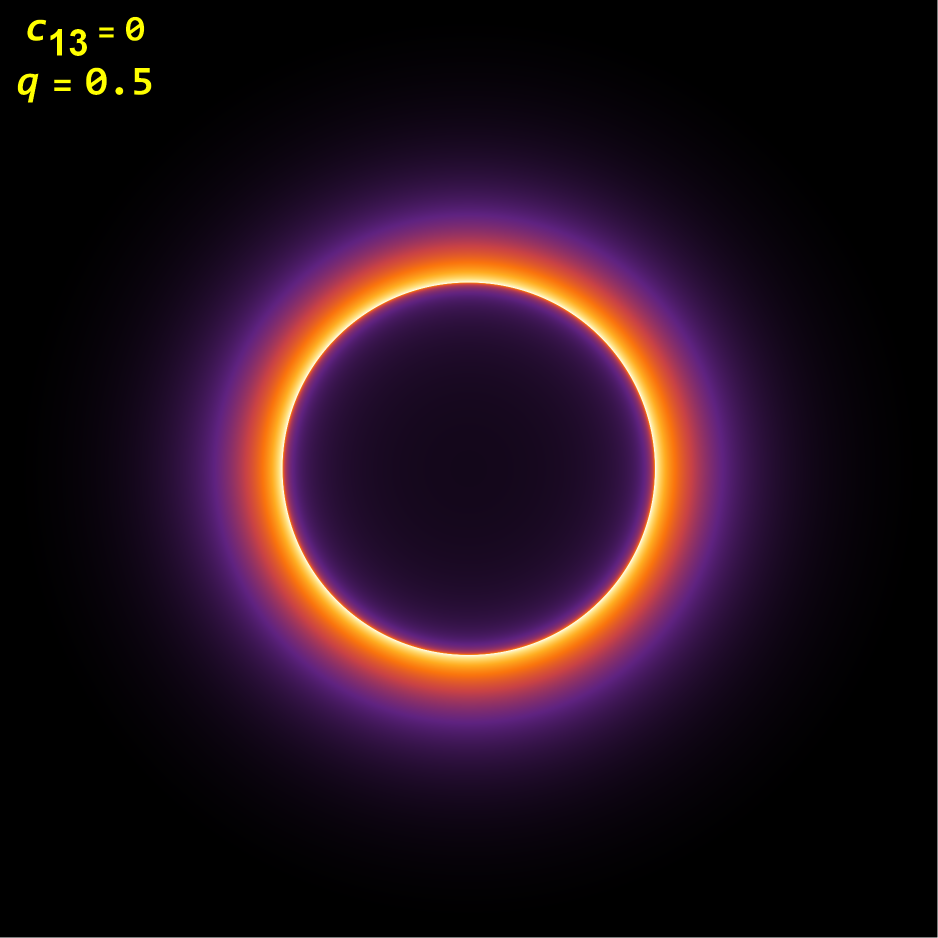}
\includegraphics[width=0.27in]{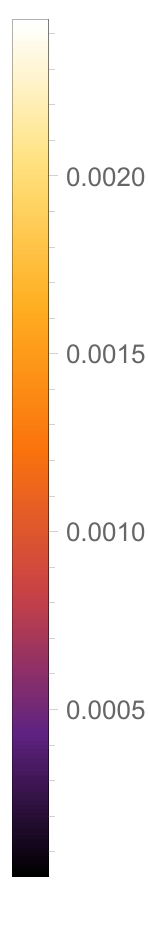}
\includegraphics[width=1.7in]{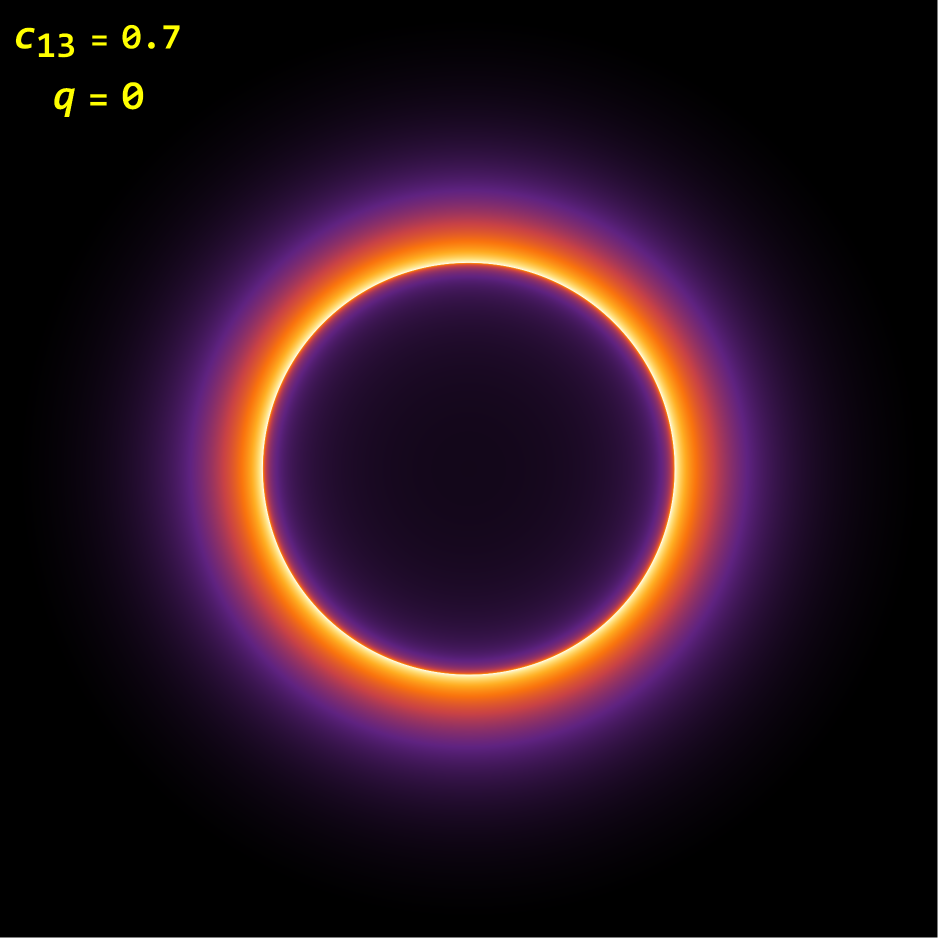}
\includegraphics[width=0.29in]{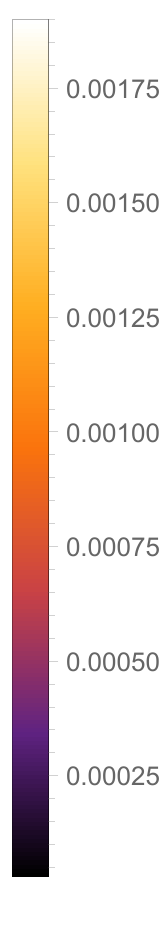}
\includegraphics[width=1.7in]{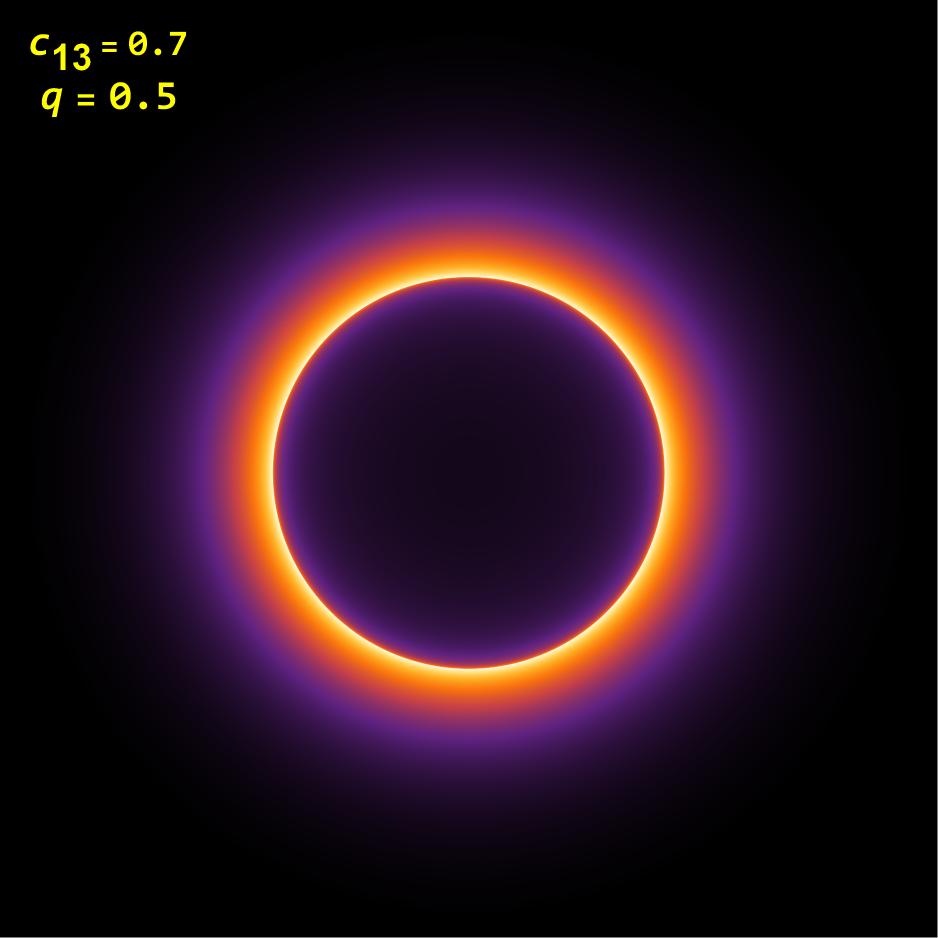}
\includegraphics[width=0.29in]{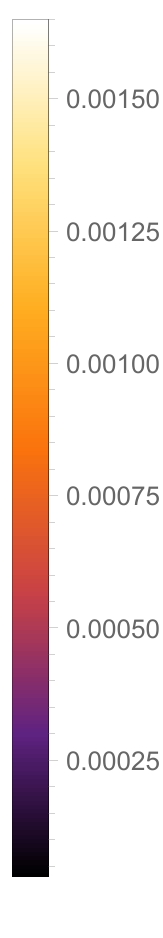}
\caption{Two dimensional image of shadows and photon rings for infalling spherical accretion flow of the RN BH with $c_{13}=0$, $q=0.5$ (left panel), the first neutral Einstein-{\AE}ther BH with $c_{13}=0.7$, $q=0$ (middle panel) and the first charged Einstein-{\AE}ther BH with $c_{13}=0.7$, $q=0.5$ (right panel) and $M = 1$.}
\label{2-dimensional-2infalling}
\end{figure}

Now, we consider charged Einstein-{\AE}ther BHs surrounded by infalling spherical accretion flow. The observed intensity in this case is still expressed by Eq.~(\ref{35}) with this difference that the redshift factor is defined as
\begin{equation}
{g^{\rm i}} = \frac{k_{\alpha}{u^{\rm i}}_{\rm ob}^{\alpha}}{k_{\beta}{u^{\rm i}}_{\rm em}^{\beta}},
\label{36}
\end{equation}
in which $k^{\mu}\equiv {\dot{x}}^{\mu}$, ${u^{\rm i}}_{\rm ob}^{\mu}=(1,0,0,0)$ and ${u^{\rm i}}_{\rm em}^{\mu}$ are the photon four-velocity, the observer four-velocity and four-velocity of the accretion flow, respectively. According to Eq.~(\ref{20}), $k_t=1/b$ is a constant and $k_r$ comes from condition $k_{\mu}k^{\mu}=0$, so we have
\begin{equation}
\frac{k_r}{k_t} = \pm\sqrt{\frac{1}{f(r)^2}-\frac{b^2}{ f(r) r^2}},
\label{37}
\end{equation}
here the sign “$\pm$” refers to the case that the photons approach $(+)$ or away $(-)$ from the charged Einstein-{\AE}ther BHs, respectively. The four-velocity of the infalling accretion gas is
\begin{equation}
({u^{\rm i}}_{\rm em}^{t}, {u^{i}}_{\rm em}^{r}, {u^{\rm i}}_{\rm em}^{\theta}, {u^{\rm i}}_{\rm em}^{\varphi} )=(\frac{1}{f(r)}, -\sqrt{1-f(r)}, 0, 0).
\label{38}
\end{equation}
Now, use of Eq.~(\ref{38}) into Eq.~(\ref{36}) leads to the following redshift factor
\begin{equation}
g^{\rm i} = \frac{1}{{u^{\rm i}}_{\rm em}^t+\left(\frac{k_r}{k_t}\right){u^{\rm i}}_{\rm em}^r}.
\label{39}
\end{equation}
Also, the infinitesimal proper distance is
\begin{equation}
dl_{\rm p} = k_{\alpha}{u^{\rm i}}^{\alpha}_{\rm em} d\tau = \frac{k_t}{{g^i} \mid k_r\mid}dr,
\label{40}
\end{equation}
where $\tau$ is the affine parameter. For a monochromatic specific emissivity, the specific intensity $I(\nu_{\rm ob})$ can be written as
\begin{equation}
I({\nu^{\rm i}}_{\rm ob}) = \int \frac{{g^{\rm i}}^2}{r^2\sqrt{\frac{1}{f(r)^2}-\frac{b^2}{f(r) r^2}}} dr.
\label{41}
\end{equation}
The observed specific intensity with respect to parameter $b$ for both the first (left panel) and the second charged Einstein-{\AE}ther BHs (right panel) is plotted in Fig.~\ref{intensity-infalling}. As can be seen, the behavior of the observed intensity is similar to the static case with this difference that the maximum value of intensity is less than that of static case. Two dimensional image of shadows and photon rings for infalling spherical accretion flow of the first charged Einstein-{\AE}ther BH has been plotted in Fig. \ref{2-dimensional-2infalling}.

\subsection{Thin disk accretion}
In this section, we aim to calculate the effect of the coupling constants and charge parameter on the observed specific intensity and luminosity of the charged Einstein-{\AE}ther BHs for thin disk accretion.

\subsubsection{The total number of light orbits}
We now consider charged Einstein-{\AE}ther BHs which is surrounded by an optically and geometrically thin disk accretion. Disk lies in equatorial plane of the BH and emits in the rest frame of the static word-line isotropically. We also consider that the static observer is at the north pole of the BH. 

According to Ref. \cite{Wald} the trajectories of light rays are divided into direct emission ($n<3/4$), lensed ring emission ($3/4<n<5/4$) and photon ring emission ($n>5/4$) which $n\equiv\frac{\varphi}{2\pi}$ is the total number of photon orbits. In direct emission the light trajectories intersect the accretion disk plane just once, but in two latter cases the light trajectories intersect the thin accretion disk twice and at least three times. For the last two cases the light rays pick up additional brightness due to the second and three intersections; so the total observed intensity is the sum of these intensities.

\begin{figure}[H]
\centering
\includegraphics[width=2.0in]{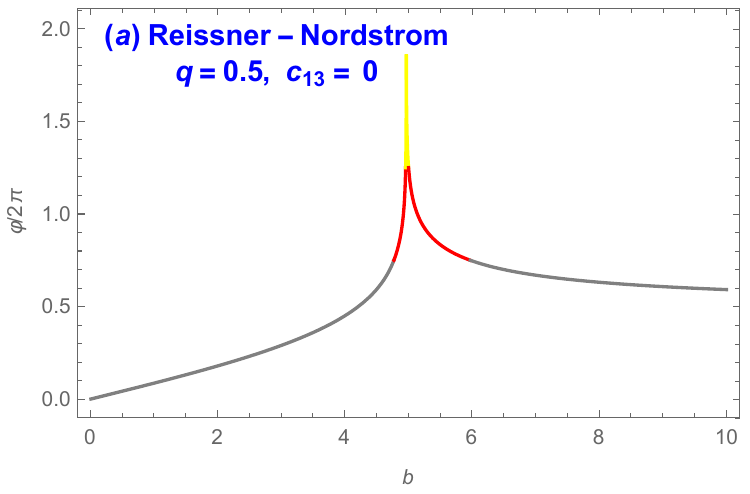}
\includegraphics[width=2.0in]{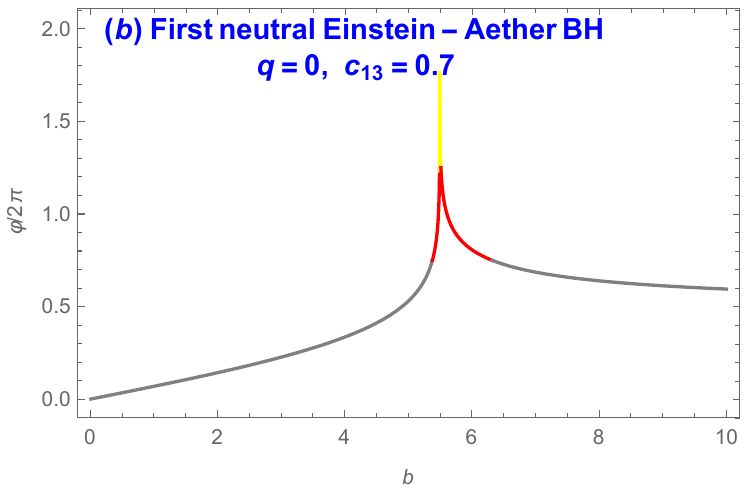}
\includegraphics[width=2.0in]{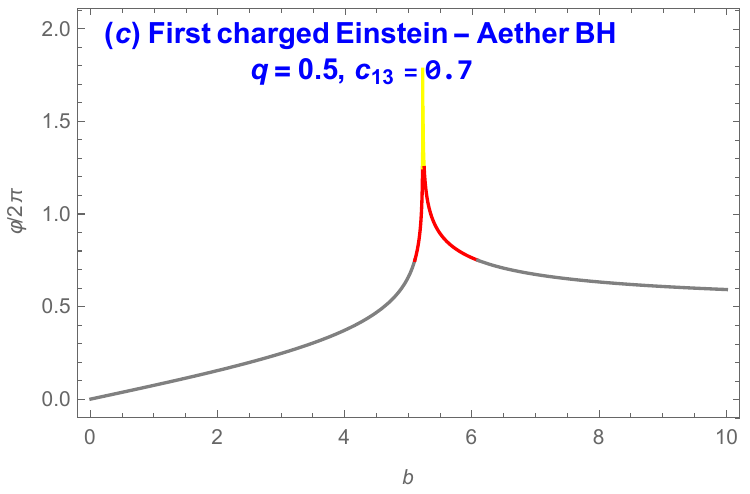}\\
\includegraphics[width=2.0in]{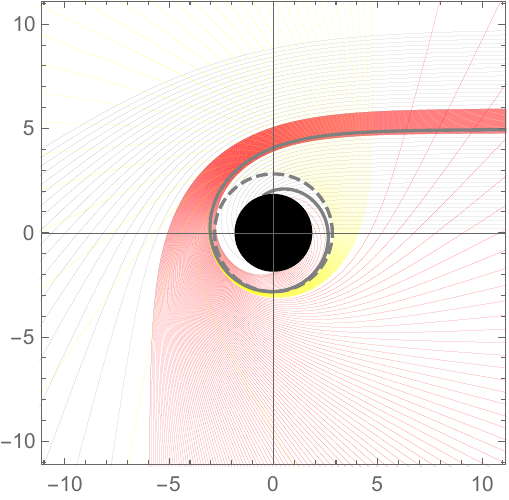}
\includegraphics[width=2.0in]{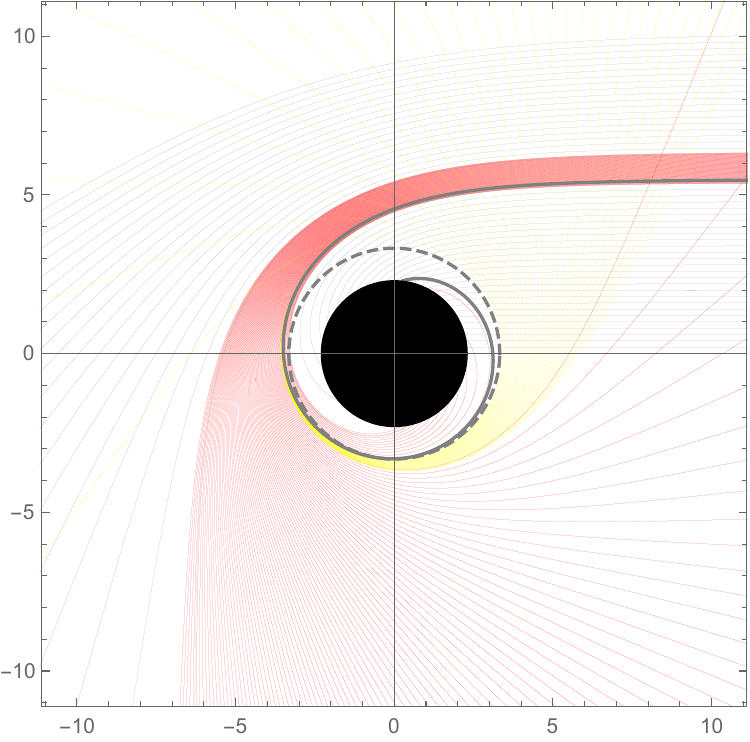}
\includegraphics[width=2.0in]{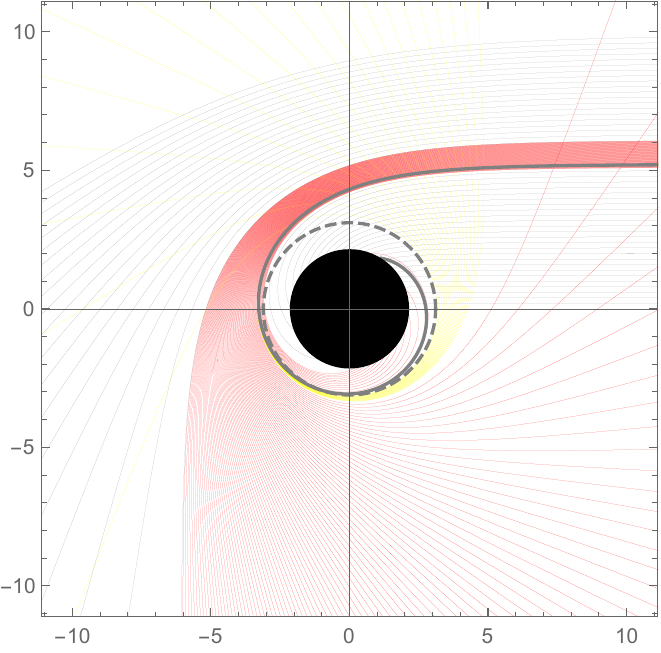}
\caption{\footnotesize The total number of photon orbits, $n$, as a function of impact parameter, $b$, for the first charged Einstein-{\AE}ther BH with $q=0.5 , c_{13}=0$ (left panel), $q=0 , c_{13}=0.7$ (middle panel) and $q=0.5, c_{13}=0.7$ (right panel), (top row). The behavior of the light trajectories in polar coordinates according to the direct emissions (gray), lensed ring (red) and photon ring (yellow), (bottom row) .}
\label{total number1}
\end{figure}

The top row of Fig. \ref {total number1} shows the total number of photon orbits as a function of impact parameter for the RN BH, the first neutral and charged Einstein-{\AE}ther BHs. As can be seen from panels (a) and (c) for a fixed value of $q$ with increasing $c_{13}$ the peak is shifted towards large $b$ and so a distant observer will see a wider internal dark region and a narrower lensed ring, whereas from panels (b) and (c) for a fixed value of $c_{13}$ with increasing $q$ the peak is shifted towards small $b$ and a distant observer will receive a reverse result. Tab. \ref{table4} shows the ranges of the impact parameter for direct, lensed ring and photon ring emissions of the first charged Einstein-{\AE}ther BH with different values of $q$ and $c_{13}$. It is easy to see that for the first charged Einstein-{\AE}ther BH with fixed value of $q$, by increasing $c_{13}$ and fixed value of $c_{13}$ with decreasing $q$, the ranges of $b$ for lensed ring and photon ring emissions get smaller. The light rays trajectories for the first charged Einstein-{\AE}ther BH are shown in the bottom row of Fig. \ref {total number1}. The central dark region represents the BH and the light rays of direct emissions, lensed ring and photon ring emissions are represented by gray, red and yellow curves, respectively.

\begin{figure}[H]
\centering
\includegraphics[width=2.0in]{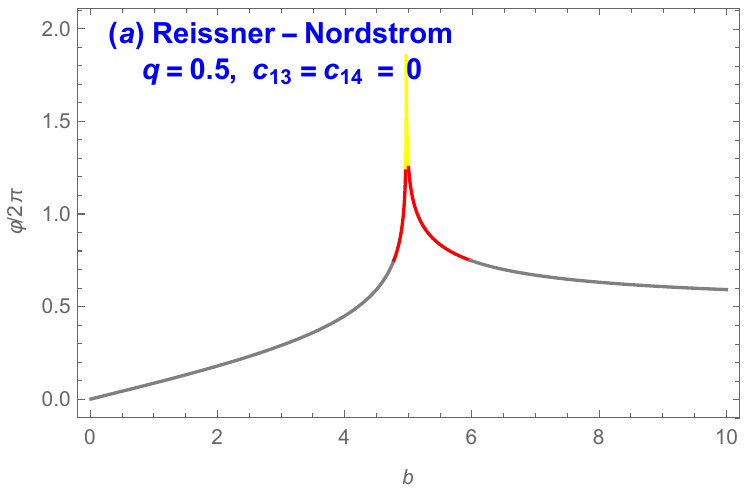}
\includegraphics[width=2.0in]{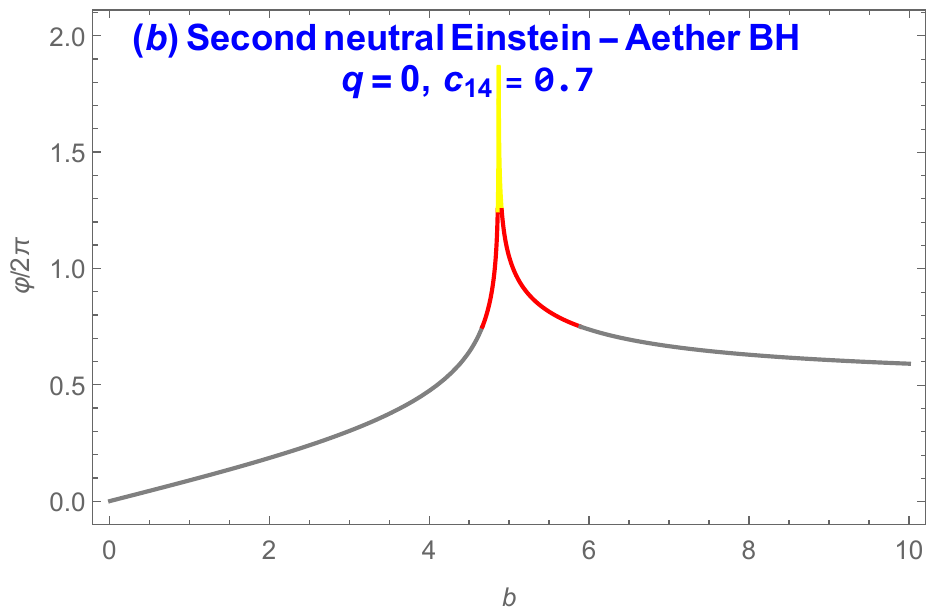}
\includegraphics[width=2.0in]{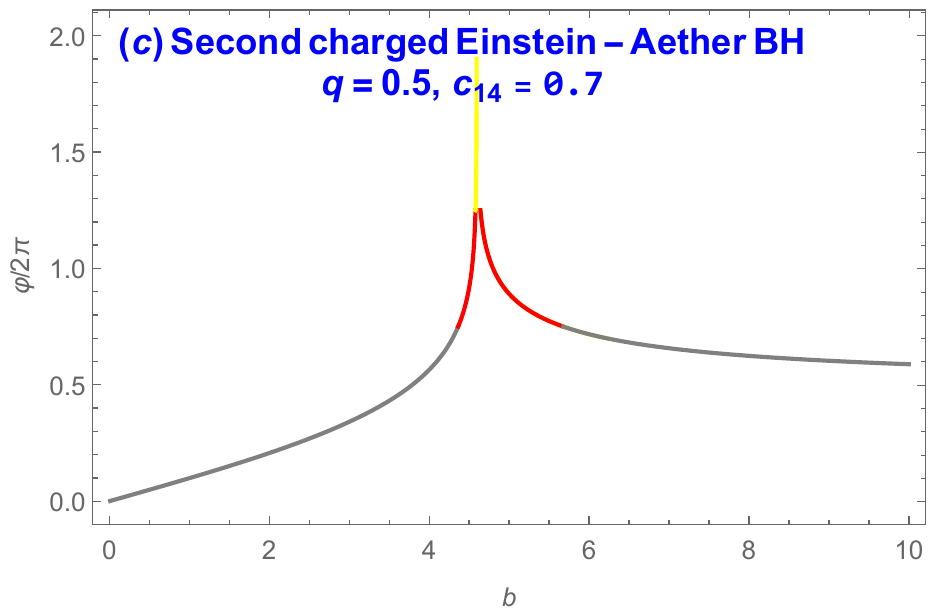}\\
\includegraphics[width=2.0in]{fig-lI50.pdf}
\includegraphics[width=2.0in]{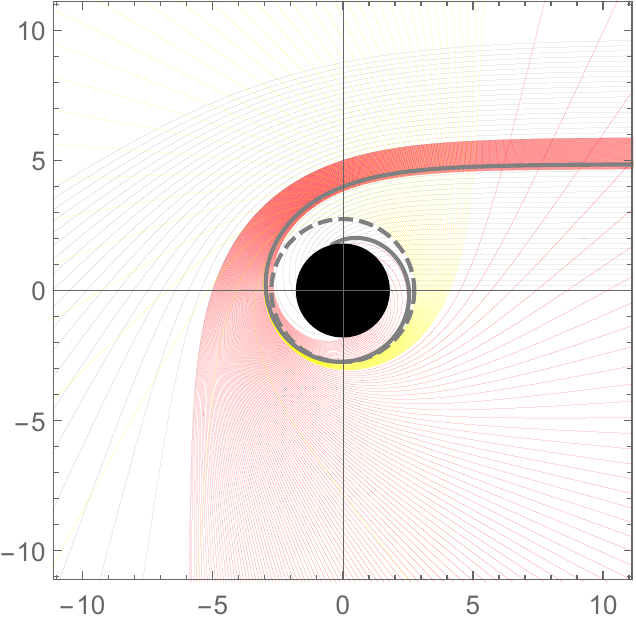}
\includegraphics[width=2.0in]{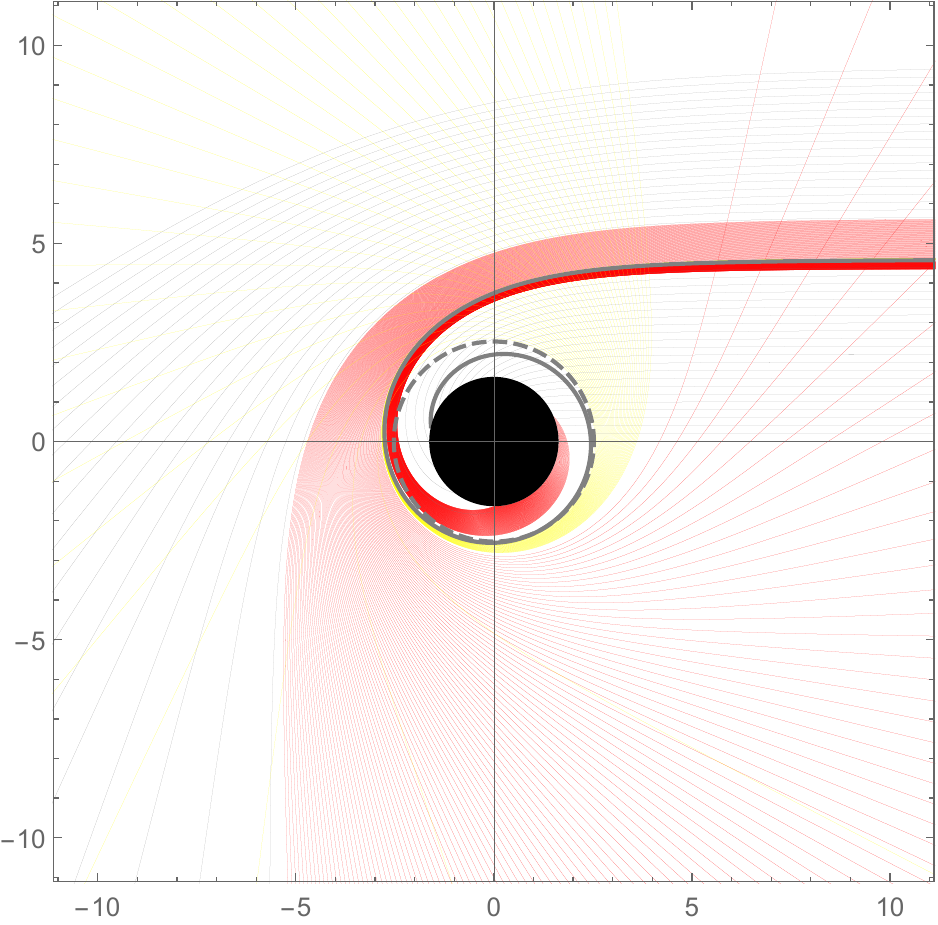}
\caption{\footnotesize The total number of photon orbits, $n$, as a function of impact parameter, $b$, for the second charged Einstein-{\AE}ther BH when $c_{13} = 0$ for $q=0.5 , c_{13}=0$ (left panel), $q=0 , c_{13}=0.7$ (middle panel) and $q=0.5, c_{13}=0.7$ (right panel), (top row). The behavior of the light trajectories in polar coordinates according to the direct emissions (gray), lensed ring (red) and photon ring (yellow), (bottom row).}
\label{total number2}
\end{figure}

Also, the top row of Fig. \ref {total number2} shows the total number of photon orbits as a function of impact parameter for the RN BH, the second neutral and charged Einstein-{\AE}ther BHs when $c_{13}=0$. Comparing panels (a) and (c) for a fixed value of $q$ with increasing $c_{14}$ and also panels (b) and (c) for a fixed value of $c_{14}$ with increasing $q$ show that the peak of curves is shifted towards small $b$ and so for both cases a distant observer will see a narrower internal dark region and a wider lensed ring. Tab. \ref{table5} shows the ranges of the impact parameter for direct, lensed ring and photon ring emissions of the second charged Einstein-{\AE}ther BHs when $c_{13} = 0$. The bottom row of Fig. \ref {total number2} shows the light rays trajectories for this case. The central dark region represents the BH and the light rays of direct emissions, lensed ring and photon ring emissions are represented by gray, red and yellow lines, respectively. In this paper, we do not have discuss the second charged Einstein-{\AE}ther BHs when $c_{14} = 0$ because the results are similar to the first type, i. e. by increasing $c_{13}$ and decreasing $q$, the peak is shifted towards large $b$.

\begin{table}[H]
\centering
\caption{\footnotesize The ranges of the impact parameter $b$ for the direct emission, lensed ring and photon ring emissions of the first charged Einstein-{\AE}ther BH with different values of $q$ and $c_{13}$.}
\begin{tabular}{|c|c|c|c|c|}
\hline
$q$&$c_{13}$&Direct emission: $n<3/4$& Lensed ring: $3/4<n<5/4$&Photon ring: $n>5/4$\\ [0.5ex]
\hline
0&  $0$ &$b<5.01514;b>6.16757$&   $5.01514<b<5.18781 ; 5.22794<b<6.16757$&   $5.19615<b<5.22794$\\
\hline
0.5&  $0.5$ &$b<4.93946; b>6.02741$&   $4.93946<b<5.0867;  5.12033<b<6.02741$ &   $5.09294<b<5.12033$\\
\hline
0.5&  $0.7$ &$b<5.10144;b>6.09346$&   $5.10144<b<5.22356 ; 5.24899<b<6.09346$&   $5.22764<b<5.24899$\\
\hline
0.7&  $0.7$&$b<4.79205; b>5.86156$&   $4.79205<b<4.92951; 4.96136<b<5.86156$ &   $4.93512<b<4.96136$\\
\hline
\end{tabular}
\label{table4}
\end{table}

\begin{table}[H]
\centering
\caption{\footnotesize The ranges of the impact parameter $b$ for the direct emission, lensed ring and photon ring emissions of the second charged Einstein-{\AE}ther BH when $c_{13} = 0$ for different values of $c_{14}$ and $q$.}
\begin{tabular}{|c|c|c|c|c|}
\hline
$q$&$c_{14}$&Direct emission: $n<3/4$& Lensed ring: $3/4<n<5/4$&Photon ring: $n>5/4$\\ [0.5ex]
\hline
0&  $0$ &$b<5.01514;b>6.16757$&   $5.01514<b<5.18781 ; 5.22794<b<6.16757$&   $5.19615<b<5.22794$\\
\hline
0.5&  $0.5$ &$b<4.49171; b>5.76663$&   $4.49171<b<4.69671;  4.75227<b<5.76663$ &   $4.70960<b<4.75227$\\
\hline
0.5& $0.7$ &$b<4.36249;   b>5.67852$&   $4.36249<b<4.57966; 4.64112<b<5.67852$ &   $4.59449<b<4.64112$\\
\hline
0.7&  $0.7$&$b<3.97657; b>5.45273$&   $3.97657<b<4.24956; 4.33793<b<5.45273$ &   $4.27475<b<4.33793$\\
\hline
\end{tabular}
\label{table5}
\end{table}

\subsubsection{The transfer function and the total observed intensity}
\begin{figure}[H]
\centering
\includegraphics[width=2.0in]{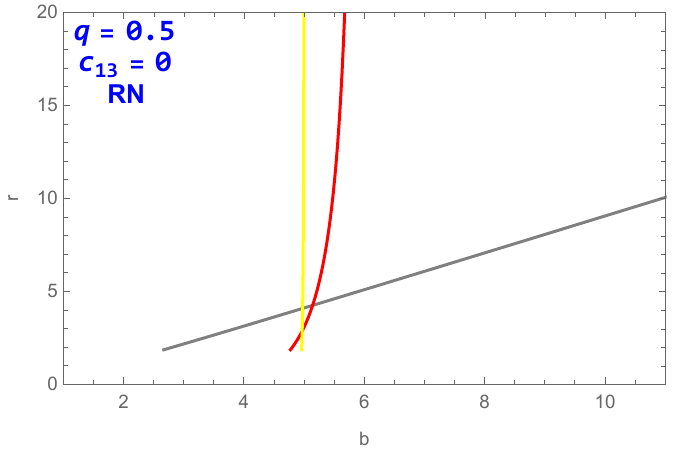}
\includegraphics[width=2.0in]{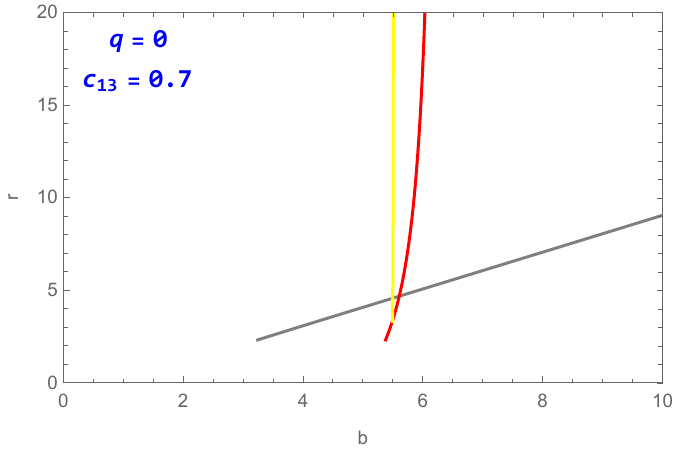}
\includegraphics[width=2.0in]{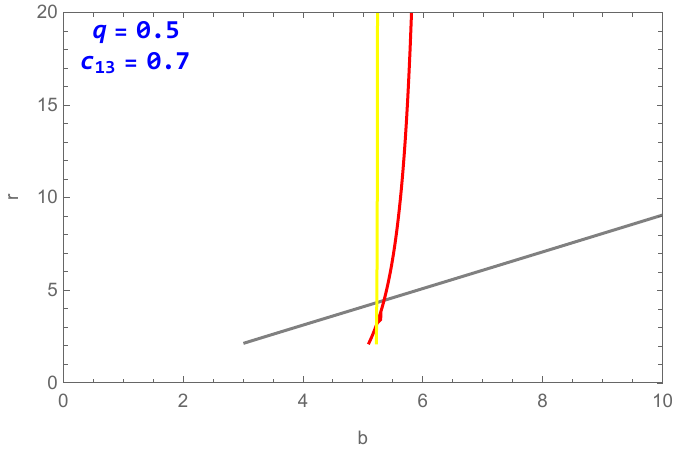}\\
\caption{\footnotesize First three transfer functions for the RN BH (left panel), first neutral Einstein-{\AE}ther BH (middle panel) and the first charged Einstein-{\AE}ther BH with $q=0.5$ and $c_{13}=0.7$ (right panel). Here the gray, red, and yellow lines correspond to the direct emission, lensed ring emission and photon ring emission, respectively.
 }
\label{transfer function1}
\end{figure}

\begin{figure}[H]
\centering
\includegraphics[width=2.0in]{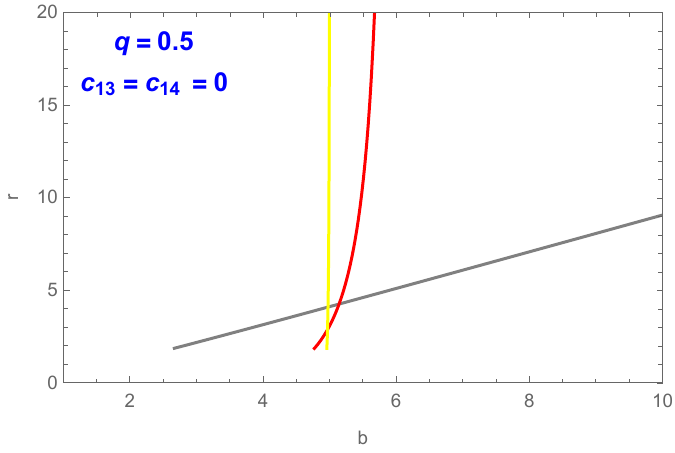}
\includegraphics[width=2.0in]{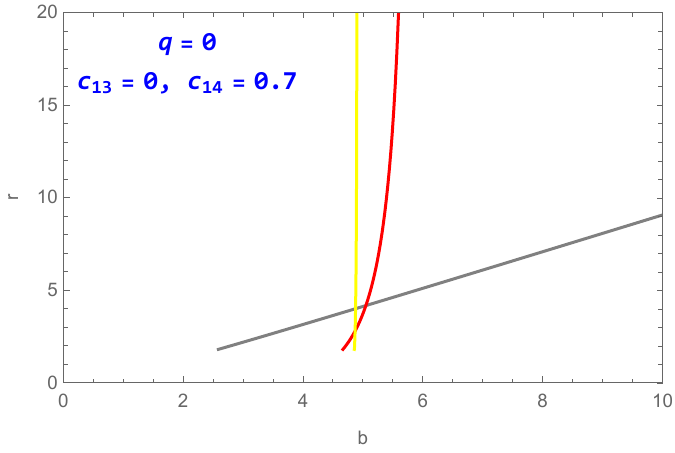}
\includegraphics[width=2.0in]{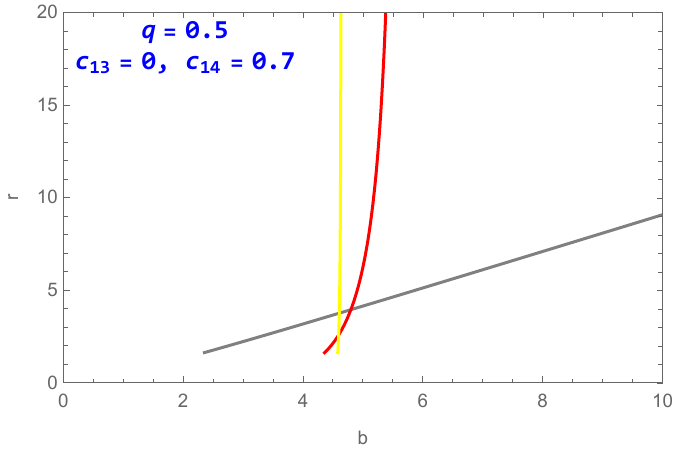}
\caption{\footnotesize First three transfer functions for the RN BH (left panel), second neutral Einstein-{\AE}ther BH (middle panel) and the second charged Einstein-{\AE}ther BHs when $c_{13} = 0$ for $c_{14}=0.7$ and $q=0.5$ (right panel). Here the gray, red, and yellow lines correspond to the direct emission, lensed ring emission and photon ring emission, respectively.
 }
\label{transfer function2}
\end{figure}

According to the Liouville's theorem, we have
\begin{equation}
\frac{I_{\rm em}(r)}{\nu_{\rm em}^3}=\frac{I_{\rm ob}(r)}{\nu_{\rm ob}^3}.
\label{42}
\end{equation}
From above equation for the charged Einstein-{\AE}ther BHs the observed intensity can be written as
\begin{equation}
I_{\rm ob}(r)=f^{\frac{3}{2}}(r)I_{\rm em}(r),
\label{43}
\end{equation}
where the observed frequency is $\nu_{\rm ob}\equiv f^{1/2}(r) \nu_{\rm em}$. By integrating the specific intensity with different frequencies we can obtain the total specific intensity for the first charged Einstein-{\AE}ther BH
\begin{equation}
I_{\rm total}(r)=\int I_{\rm ob}(r)d\nu_{\rm ob} = \left[1-\frac{2M}{r}+\frac{q^{2}}{r^{2}}+\frac{c_{13}B}{\left(1-c_{13}\right)r^{4}}\right]^{2} I_{\rm emit}(r),
\label{44}
\end{equation}
and for the second charged Einstein-{\AE}ther BH as
\begin{equation}
I_{\rm total}(r)=\int I_{\rm ob}(r)d\nu_{\rm ob} = \left[1-\frac{2M}{r}-\frac{r_{u}\left(2M+r_{u}\right)}{r^{2}}\right]^{2} I_{\rm emit}(r),
\label{44}
\end{equation}
where
\begin{equation}
I_{\rm emit}(r)=\int I_{\rm em}(r)d\nu_{\rm em},
\label{45}
\end{equation}
is the total emitted specific intensity of accretion disk. As we mentioned above, due to intersections of light rays with the disk plane, the total observed intensity is the sum  of  extra intensities from each intersection, that is
\begin{equation}
I_{\rm total}(r)=\sum_{n}f^{2}(r)I_{\rm emit}(r)|_{r=r_{\rm n}(b)},
\label{46}
\end{equation}
where $r_{\rm n}(b)$ is the transfer function that indicates the radial position of the $n_{\rm th}$ intersection between the light rays with a given impact parameter and the thin accretion disk \cite{Wald}. We have plotted the first three transfer functions for the RN BH (left panel), first neutral Einstein-{\AE}ther BH (middle panel) and the first charged Einstein-{\AE}ther BH (right panel) in Fig. \ref{transfer function1}. Also, the first three transfer functions for the RN BH (left panel), second neutral Einstein-{\AE}ther BHs (middle panel) and for the second charged Einstein-{\AE}ther BHs (right panel) when $c_{13} = 0$ have been plotted in Fig. \ref{transfer function2}. The first transfer function with $n=1$ described by the gray lines corresponds to “direct emission”. The second and third transfer function with $n=2$ and $n=3$ represented by the red and yellow lines correspond to “lensed ring” and “photon ring”, respectively.

Now, we aim to investigate the optical appearances of the charged Einstein-{\AE}ther BHs. It was found that the distribution of disk radiation is nearly a Gaussian function \cite{G1, G2}, so we consider the radiation intensity as a Gaussian function
\begin{equation}
        I_{\rm em}(r)=
        \begin{cases}
            \exp\left[\frac{-(r-r_{\rm in})^2}{8}\right] &r>r_{\rm in}, \\
            0 &r\leq r_{\rm in},
        \end{cases}\label{31}
    \end{equation}
where $r_{\rm in}$  is the innermost radiation radius of the thin accretion disk. In what follows, we select three different scenarios: case (A) $r_{\rm in}=r_{\rm isco}$, where $r_{\rm isco}$ is the radius of the Innermost Stable Circular Orbit (ISCO) of the BH, case (B) $r_{\rm in}=r_{\rm ph}$ in which $r_{\rm ph}$ is the photon sphere radius of the BH  and case (C)$r_{\rm in}=r_{h}$ where $r_{\rm h}$ is the BH event horizon radius.

$\bullet $ Case A: $r_{\rm in}=r_{\rm isco}$

\begin{figure}[H]
\centering
\includegraphics[width=2.0in]{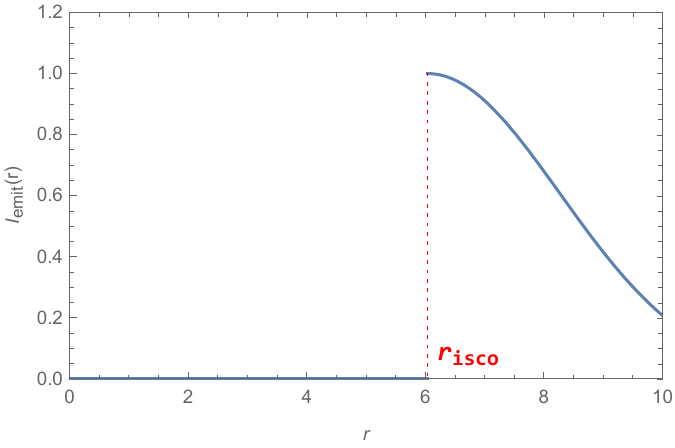}
\includegraphics[width=2.0in]{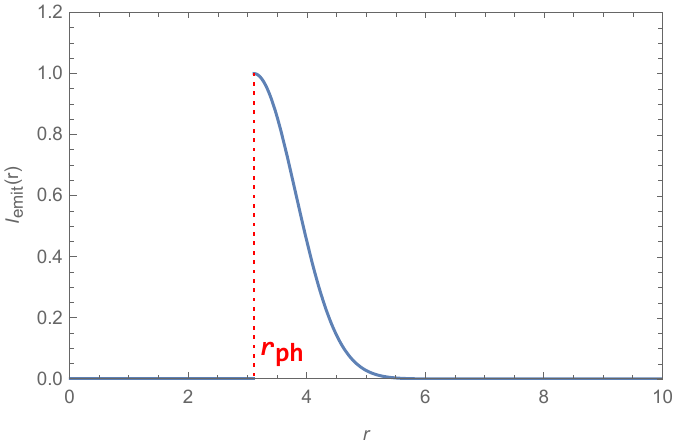}
\includegraphics[width=2.0in]{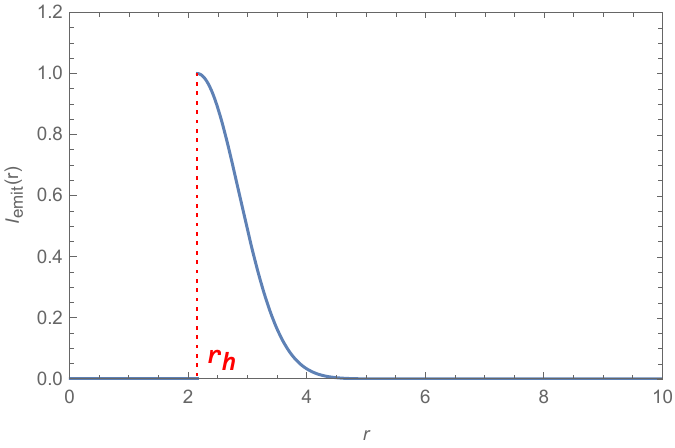}\\
\includegraphics[width=2.0in]{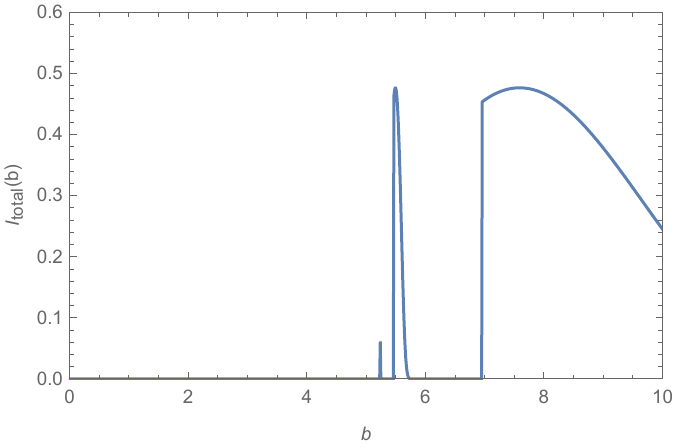}
\includegraphics[width=2.0in]{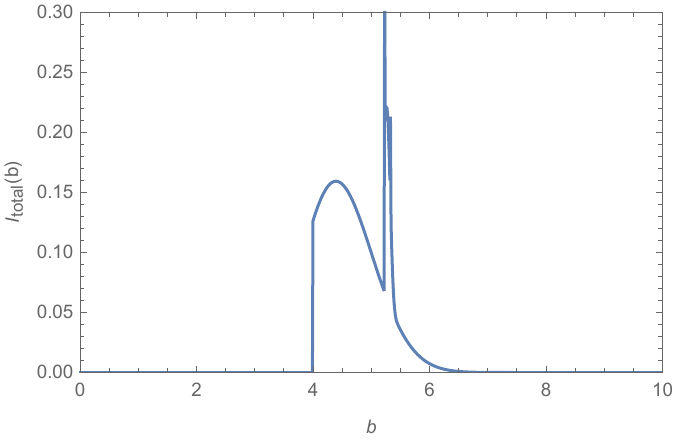}
\includegraphics[width=2.0in]{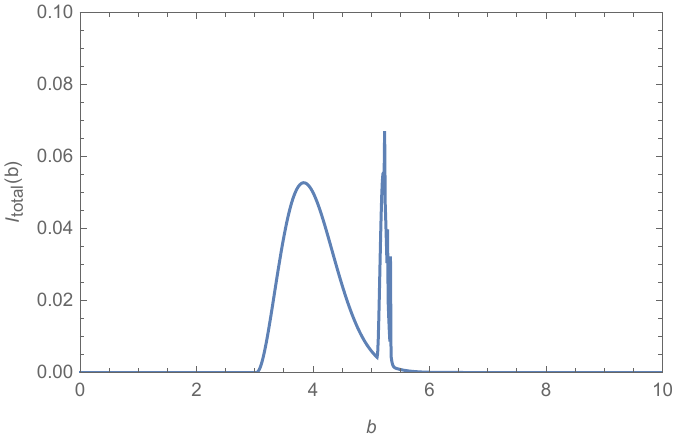}\\
\includegraphics[width=1.98in]{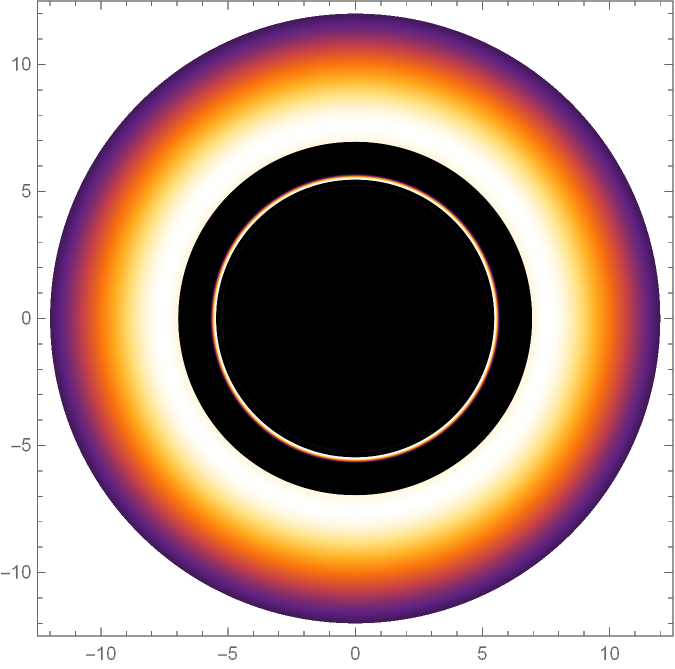}
\includegraphics[width=1.98in]{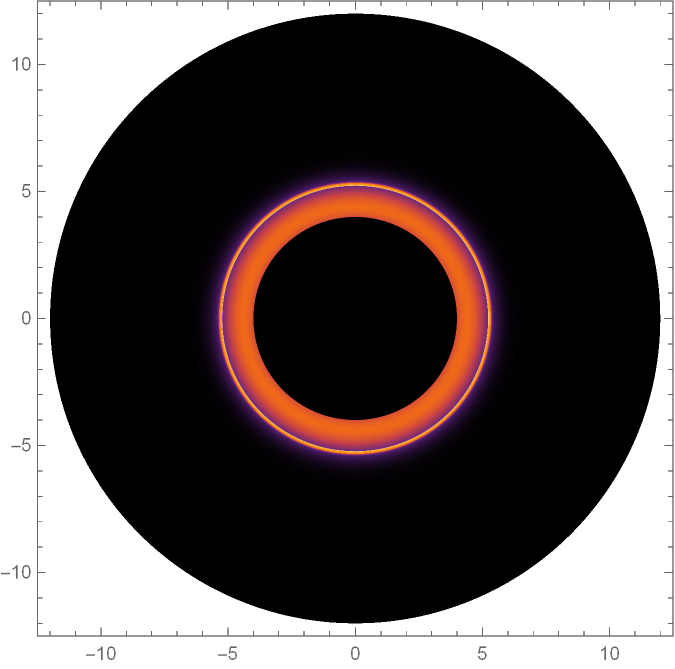}
\includegraphics[width=1.98in]{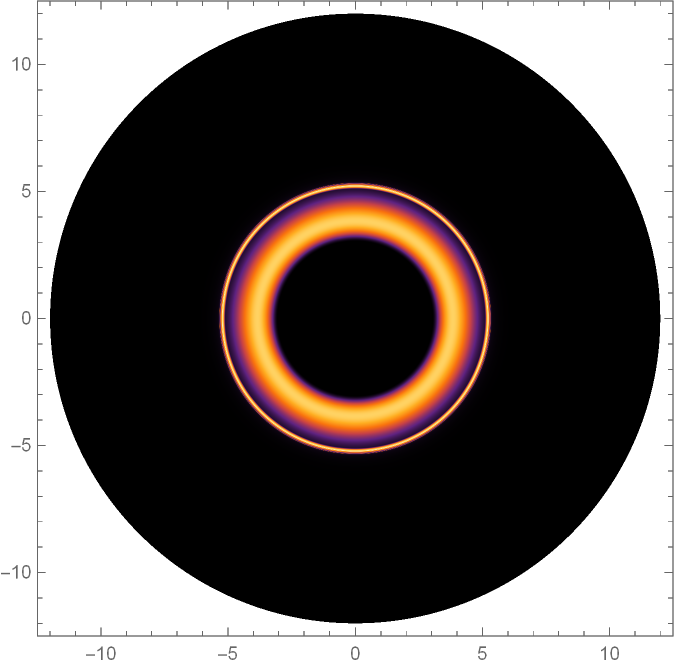}\\
\caption{\footnotesize The total radiation intensity as a function of radius (top row), the total observed intensity as a function of the impact parameter (middle row) and the two dimensional images of the shadows for the first charged Einstein-{\AE}ther BH (bottom row) with $c_{13}=0.7$, $q=0.5$ and $M=1$. }
\label{intensity1}
\end{figure}

ISCO is the last circular orbit which it has a minimal radius for a particle orbiting around the BH. A particle will plunge into the BH when its
orbit radius is less than of the ISCO. It can be obtained from the following equation
\begin{equation}
r_{\rm isco}=\frac{3f(r_{\rm isco})f^{'}(r_{\rm isco})}{2{f^{'}}^{2}(r_{\rm isco})-f(r_{\rm isco})f^{''}(r_{\rm isco})}.
\label{48}
\end{equation}
Let us focus on the first charged Einstein-{\AE}ther BH with $c_{13}=0.7$, $q=0.5$ and $r_{\rm isco} = 6.03659$. In the left column of Fig. \ref{intensity1}, the total radiation intensity as a function of $r$, the total observed intensity as a function of $b$ and two-dimensional image are presented. The direct emission starts at $b\simeq 6.96$ and peaks at $b\simeq 7.59$ with $I_{\rm max}\simeq 0.47668$. The limit of the lensed ring is in a small range $b\simeq 5.47 \sim 5.76$ and the photon ring starts at $b \simeq 5.24$. For the first neutral Einstein-{\AE}ther BH with $c_{13}=0.7$, $q=0$ and $r_{\rm isco}=6.50364$ the positions of the three peaks are $b \simeq 8.02$, $5.75$ and $5.51$. Comparing the first charged Einstein-{\AE}ther BH to first neutral Einstein-{\AE}ther BH shows that by increasing the charge parameter the position of the three peaks moves to the left. Three peaks corresponding to the direct emission, lensed ring and photon ring in RN BH are $b\simeq 7.23, 5.31$ and $4.99$. Also, by comparing the first charged Einstein-{\AE}ther BH to RN BH shows that in the presence of {\AE}ther field the position of the three peaks move to the right. We summarized the values of the position of the three peaks for Schwarzschild BH, RN BH, first neutral Einstein-{\AE}ther BH and the first charged Einstein-{\AE}ther BH in Tab. \ref{table}.

\begin{figure}[H]
\centering
\includegraphics[width=2.0in]{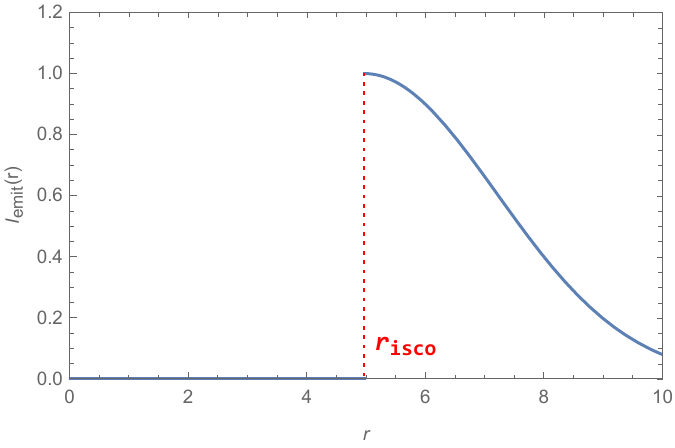}
\includegraphics[width=2.0in]{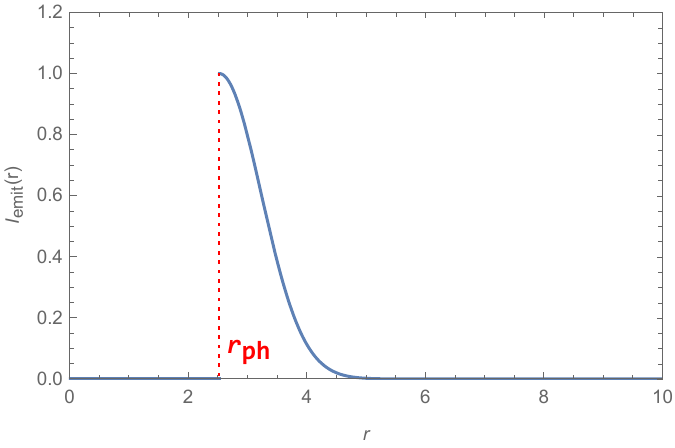}
\includegraphics[width=2.0in]{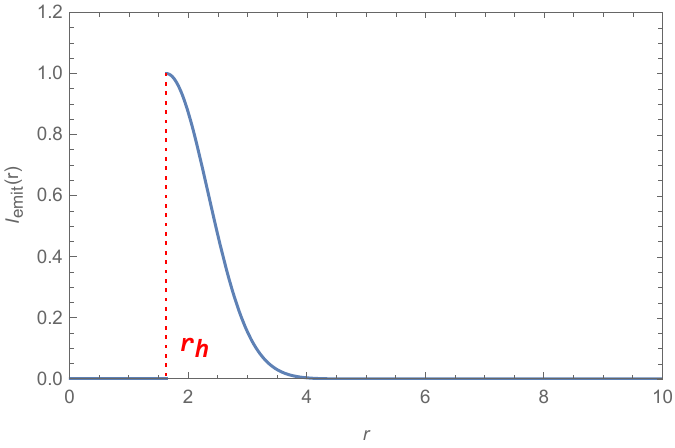}\\
\includegraphics[width=2.0in]{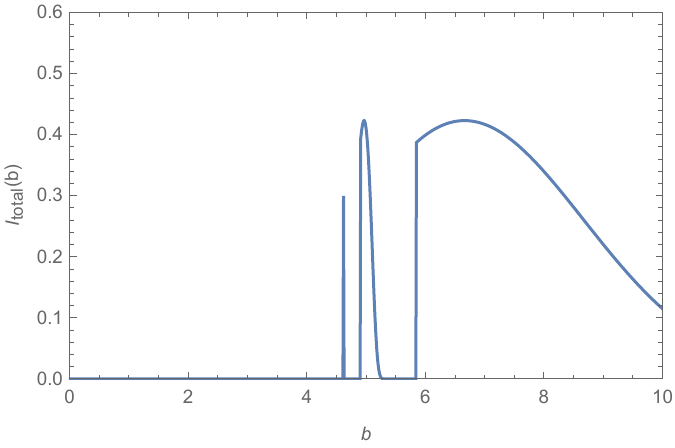}
\includegraphics[width=2.0in]{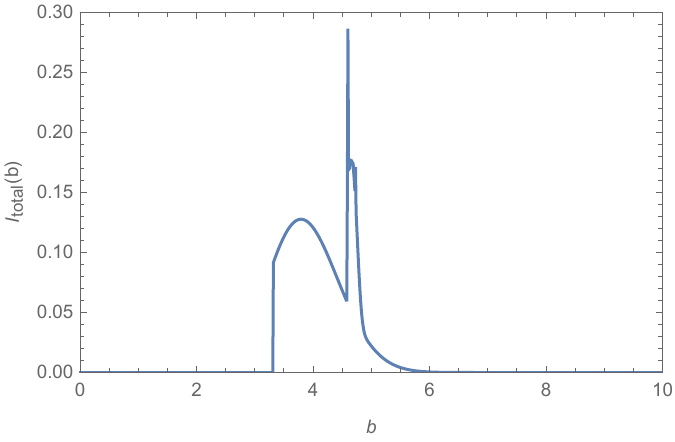}
\includegraphics[width=2.0in]{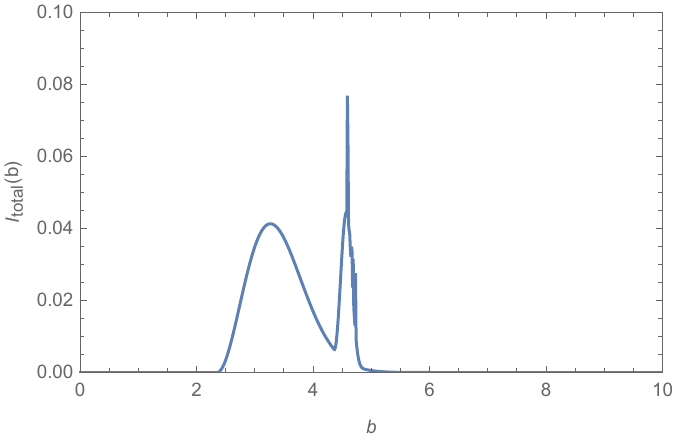}\\
\includegraphics[width=1.98in]{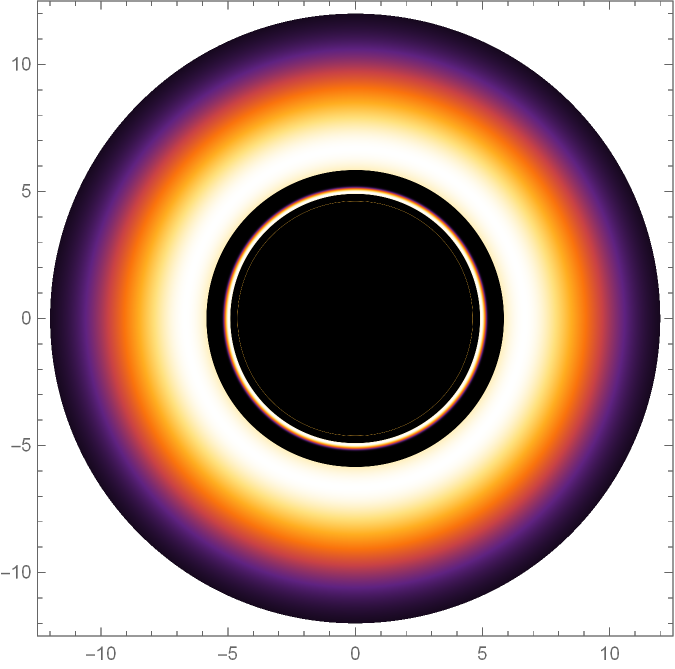}
\includegraphics[width=1.98in]{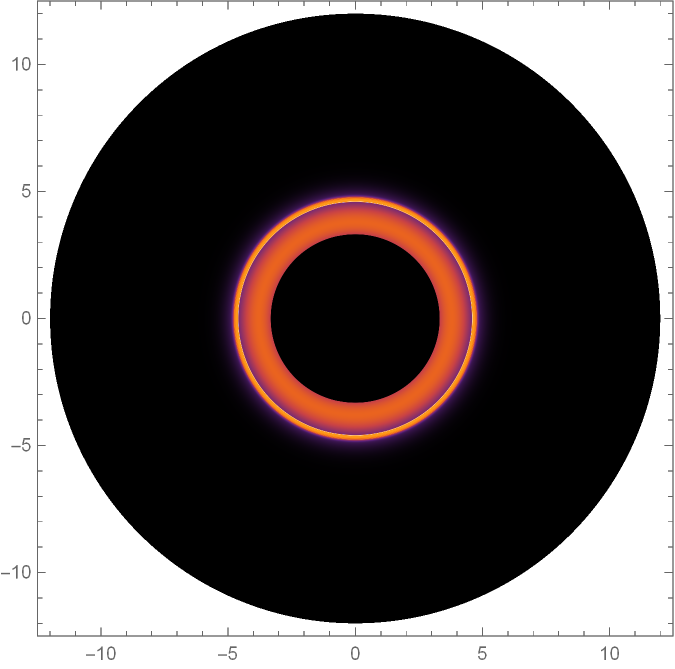}
\includegraphics[width=1.98in]{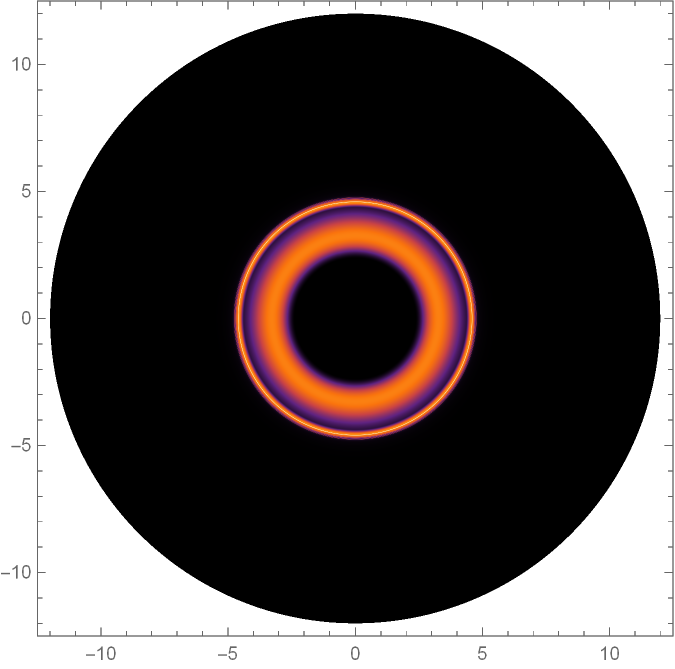}\\
\caption{\footnotesize The total radiation intensity as a function of radius (top row), the total observed intensity as a function of the impact parameter (middle row) and the two dimensional images of the shadows for the second charged Einstein-{\AE}ther BHs when $c_{13}=0$ (bottom row) for $c_{14}=0.7$ and $q=0.5$ with $M=1$. }
\label{intensity2}
\end{figure}

For the second charged Einstein-{\AE}ther BH when $c_{13}=0$ with $c_{14}=0.7$ and $q=0.5$ the total radiation intensity as a function of $r$, the total observed intensity as a function of $b$ and two-dimensional image are plotted in the left column of Fig. \ref{intensity2}. The direct emission starts at $b\simeq 5.85$ and peaks at $b\simeq 6.66$ with $I_{\rm max}\simeq 0.42298$. The limit of the lensed ring is in a small range $b\simeq 4.91 \sim 5.31$ and the photon ring starts at $b \simeq 4.62$. Also, for the second neutral Einstein-{\AE}ther BHs when $c_{13}=0$ with $c_{14}=0.7$ the positions of the three peaks are $b \simeq 7.07$, $5.21$ and $4.89$. Therefore for the second type by increasing both the charge parameter and the coupling constant the position of the three peaks moves to the left.

$\bullet $ Case B: $r_{\rm in}=r_{\rm ph}$

In this case for the first charged Einstein-{\AE}ther BH with $c_{13}=0.7$ and $q=0.5$, we obtain $r_{\rm ph} = 3.10872$. In the middle column of Fig. \ref{intensity1}, the total radiation intensity as a function of $r$, the total observed intensity as a function of $b$ and two-dimensional image are presented. The direct emission starts at $b\simeq 4.01$ and peaks at $b\simeq 4.39$ with $I_{\rm max}\simeq 0.15939$. The limit of the lensed ring is in a small range $b\simeq 5.23 \sim 5.48$ and the photon ring starts at $b \simeq 5.23$. The corresponding results for the second charged Einstein-{\AE}ther BH when $c_{13} = 0$ with $c_{14} = 0.7$ and $q = 0.5$ are also plotted in the middle column of Fig. \ref{intensity2}. The direct emission starts at $b\simeq 3.32$ and peaks at $b\simeq 3.80$ with $I_{\rm max}\simeq 0.12777$. The limit of the lensed ring is in a small range $b\simeq 4.59 \sim 4.96$ and the photon ring starts at $b \simeq 4.60$.

$\bullet $ Case C: $r_{\rm in}=r_{\rm H}$

For third case when innermost of the accretion disk position is in the BH event horizon, the radiation peak for the first charged Einstein-{\AE}ther BH with $c_{13}=0.7$ and $q=0.5$ is in $r_{\rm h}=2.15403$. The results are plotted in the right column of Fig. \ref{intensity1}. The direct emission starts at $b\simeq 3.04$ and peaks at $b\simeq 3.84$ with $I_{\rm max}\simeq 0.05271$. The limit of the lensed ring is in a small range $b\simeq 5.11 \sim 5.42$ and the photon ring starts at $b \simeq 5.23$. For the second charged Einstein-{\AE}ther BH when $c_{13}=0$ with $c_{14}=0.7$ and $q=0.5$ the total radiation intensity as a function of $r$, the total observed intensity as a function of $b$ and two-dimensional image are plotted in the right column of Fig. \ref{intensity2}. The direct emission starts at $b\simeq 2.37$ and peaks at $b\simeq 3.27$ with $I_{max}\simeq 0.04132$. The limit of the lensed ring is in a small range $b\simeq 4.37 \sim 4.89$ and the photon ring starts at $b \simeq 4.58$.

\begin{table}[H]
\centering
\caption{\footnotesize The position of the three peaks corresponding to direct emission, lensed ring and photon ring, for Schwarzschild BH, RN BH, first neutral Einstein-{\AE}ther BH and the first charged Einstein-{\AE}ther BH.}
\begin{tabular}{|c|c|c|c|c|}
\hline
Type&ISCO radius&Direct emission& Lensed ring&Photon ring\\ [0.5ex]
\hline
Schwarzschild&  $6.0$ &$7.0$&   $5.5$&   $5.2$\\
\hline
Reissner-Nordstrom&  $5.60664$ &$7.23$&   $5.31$&   $4.99$\\
\hline
First neutral Einstein-{\AE}ther BH& $6.50364$ &$8.02$&   $5.75$ &   $5.51$\\
\hline
First charged Einstein-{\AE}ther BH&  $6.03659$ &$7.59$&   $5.5$ &   $5.24$\\
\hline
\end{tabular}
\label{table}
\end{table}

\section{Conclusions\label{section6}}
In this paper, we briefly reviewed the Einstein-Maxwell-{\AE}ther theory and introduced two types of the charged Einstein-{\AE}ther BHs including the  first type of charged Einstein-{\AE}ther BH with $c_{123} \neq 0$ and the second type of charged Einstein-{\AE}ther BHs with $c_{123} = 0$; when $c_{14} = 0$ and when $c_{13} = 0$ \cite{Ding:2015kba}. The effects of the coupling constants on the optical appearance of the Einstein-{\AE}ther BHs have been discussed in Ref. \cite{Aether}. Here we investigated the effects of both the coupling constants and charge parameter on the BH event horizon $r_{\rm h}$, photon sphere $r_{\rm ph}$ and impact parameter $b_{\rm ph}$ of these BHs. We found that the increase of charge leads to the decrease of these BH parameters in both the first and the second charged Einstein-{\AE}ther BHs. The results were summarized in Tab. \ref{table1}, \ref{table2} and \ref{table3}. So, the presence of the charge parameter in two types of Einstein-{\AE}ther BHs caused that the BH shadow radius and photon rings is shrunk inward the charged Einstein-{\AE}ther BHs.

Then, we studied the charged Einstein-{\AE}ther BHs surrounded by static/infalling accretion flows and calculated the observed specific intensity. Figs.~ \ref{intensity-static} and \ref{intensity-infalling} show the peak of intensity decreases by increasing $c_{13}$ for the first charged Einstein-{\AE}ther BH and the second charged Einstein-{\AE}ther BH when $c_{14} = 0$, whereas for the second charged Einstein-{\AE}ther BH when $c_{13} = 0$ the peak of intensity increase by increasing $c_{14}$. For two types of BHs the charge parameter shrunk the BH shadow radius inward the BH and so increases the peak of intensity in comparison to the neutral Einstein-{\AE}ther BH. The results also show that for an infalling accretion, due to the Doppler effect of infalling matter, the dark central region is darker than the static case. Two dimensional image of shadow and photon rings have been plotted in Figs. \ref{2-dimensional-2static} and \ref{2-dimensional-2infalling}.

For the thin disk accretion model, following the method proposed by Gralla et al. \cite{Wald}, we studied the photon trajectories with different total number of orbits. The behavior of the photon for the first charged Einstein-{\AE}ther BH and the second charged Einstein-{\AE}ther BH when $c_{14} = 0$ is shown in Fig. \ref{total number1} and Fig. \ref{total number2}, respectively. The total observed intensity as a function of the impact parameter has been plotted in Fig. \ref{intensity1} and Fig. \ref{intensity2} for the first and second charged Einstein-{\AE}ther BHs. We found that in the thin disk accretion flow the location and the emitted model of the accretion gas affect on the optical appearance of charged Einstein-{\AE}ther BHs in contrast to the spherical accretion flows. Moreover, the results indicate that in all three emitted models the direct emission has the dominant contribution to the optical appearance of the BH image.

\end{document}